\documentclass[twocolumn,prl]{revtex4}
\usepackage[latin9]{inputenc}
\usepackage{color}
\usepackage{verbatim}
\usepackage{amsmath}
\usepackage{amssymb}
\usepackage{graphicx}
\usepackage{esint}
\usepackage{hyperref}

\makeatletter
\@ifundefined{textcolor}{}
{%
 \definecolor{BLACK}{gray}{0}
 \definecolor{WHITE}{gray}{1}
 \definecolor{RED}{rgb}{1,0,0}
 \definecolor{GREEN}{rgb}{0,1,0}
 \definecolor{BLUE}{rgb}{0,0,1}
 \definecolor{CYAN}{cmyk}{1,0,0,0}
 \definecolor{MAGENTA}{cmyk}{0,1,0,0}
 \definecolor{YELLOW}{cmyk}{0,0,1,0}
}

\usepackage{mathrsfs}

\draft

\makeatother

\begin{document}
\title{From semimetal to chiral Fulde-Ferrell superfluids}
\author{Ting Fung Jeffrey Poon}
\affiliation{International Center for Quantum Materials, School of Physics, Peking University, Beijing 100871, China}
\affiliation{Collaborative Innovation Center of Quantum Matter, Beijing 100871, China}
\author{Xiong-Jun Liu
\footnote{Corresponding author: xiongjunliu@pku.edu.cn}}
\affiliation{International Center for Quantum Materials, School of Physics, Peking University, Beijing 100871, China}
\affiliation{Collaborative Innovation Center of Quantum Matter, Beijing 100871, China}
\begin{abstract}
The recent realization of two-dimensional (2D) synthetic spin-orbit (SO) coupling opens a broad avenue to study novel topological states for ultracold atoms. Here, we propose a new scheme to realize exotic chiral Fulde-Ferrell superfluid for ultracold fermions, with a generic theory being shown that the topology of superfluid pairing phases can be determined from the normal states. The main findings are two fold. First, a  semimetal is driven by a new type of 2D SO coupling whose realization is even simpler than the recent experiment, and can be tuned into massive Dirac fermion phases with or without inversion symmetry. Without inversion symmetry the superfluid phase with nonzero pairing momentum is favored under an attractive interaction. Furthermore, we show a fundamental theorem that the topology of a 2D chiral superfluid can be uniquely determined from the unpaired normal states, with which the topological chiral Fulde-Ferrell superfluid with a broad topological region is predicted for the present system. This generic theorem is also useful for condensed matter physics and material science in search for new topological superconductors.
\end{abstract}

\maketitle
\indent
\textit{Introduction.}--The recent experimental realization of two-dimensional (2D) spin-orbit (SO) coupling for ultracold atoms~\cite{Zhan2016,Zhang2016a,Zhang2016b}, which corresponds to synthetic non-Abelian gauge potentials~\cite{Ruseckas2005,Zoller2005,Liu2009}, advances an essential step to explore novel topological quantum phases beyond natural conditions. Ultracold fermions with SO coupling can favor the realization of topological superfluids (TSFs) (similar as topological superconductors in solids~\cite{Read2000,Kitaev2001,Kitaev2003,Fu2008,Sau2010,Lutchyn2010,Oreg2010}) based on an $s$-wave Feshbach resonance~\cite{Zhang2008,Sato2009}, which are highly-sought-after quantum phases for their ability to host non-Abelian Majorana zero modes and implement topological quantum computation~\cite{Wilczek2009,Alicea2012,Franz2013,Elliott2015}. Note that a superfluid phase has to exist in 2D or 3D regime, so having a 2D or 3D SO coupling is the basic requirement for such realization of gapped TSFs. While experimental studies of TSFs are yet to be available, different proposals have been introduced for Rashba and Dirac type SO coupled systems~\cite{Zhu2011,Chunlei2013,Zhang2013,Hu2014,Liu2014,Chan2016}, with BCS or FFLO pairing orders~\cite{FF1964,LO1964}. When the Fermi energy crosses only a single (or odd number of) Fermi surface (FS), the SO coupling forces Cooper pairs into effective $p$-wave type, rendering a TSF phase~\cite{Alicea2012}. However, in the generic case it is not known so far whether there is a universal way to precisely determine the topology, e.g. Chern numbers, of a superfluid phase induced in normal bands. On the other hand, to achieve a TSF phase integrates several essential ingredients, which may bring challenges for the experiment. The minimal schemes of realization are therefore desired to ensure high feasibility of real studies.

In this letter we propose an experimental scheme to realize chiral Fulde-Ferrell (FF) superfluids, and show a generic theory to compute the Chern number of TSF phases through the normal states. The findings are of significance in both experiment and fundamental theory. First, we propose that a new type of 2D SO coupled Dirac semimetal can be realized based on a scheme adopting only a single Raman transition and simpler than the recent experiments~\cite{Zhan2016,Zhang2016a,Zhang2016b}. By readily breaking the inversion symmetry of the Dirac system, we find that the FF superfluid phase can be favored under an attractive interaction. Moreover, we show a generic formalism for the Chern number of a 2D superfluid induced in the normal states, with which the topological chiral FF superfluid with a broad topological region is predicted.

\begin{figure}[h]
\centerline{\includegraphics[width=1\columnwidth]{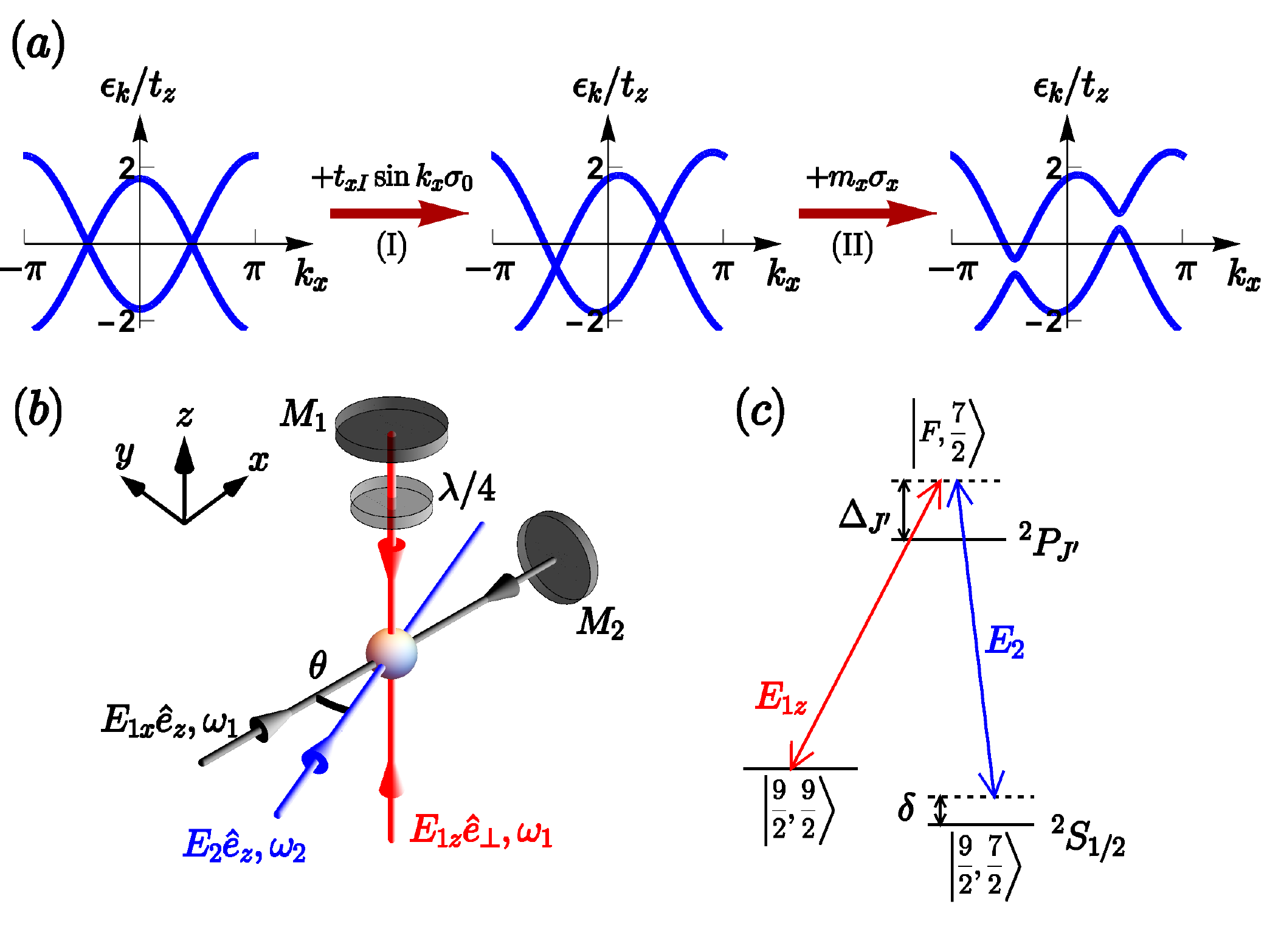}}
\caption{(a) Schematic diagram of breaking the inversion symmetry (process I) and opening gap at Dirac points (II) for a 2D SO coupled Dirac semimetal. (b) Proposed experimental setting for realization. The standing wave lights formed by $\bold E_{1x,1z}$ generate a blue-detuned square lattice. The incident polarization of $\bold E_{1z}$ is $\hat{e}_\bot = \alpha \hat e_x + i\beta \hat e_y$, and the $\lambda/4$-wave plate changes the polarization of the reflected field to $\hat{e}'_\bot = \alpha \hat e_x - i\beta \hat e_y$. The Raman coupling, illustrated in (c) for $^{40}$K fermions, is generated by $\bold E_{1z}$ and an additional running light $\bold E_2$ which has tile angle $\theta$ with respect to $x$-$z$ plane.}
\label{Fig:Realization}
\end{figure}

\textit{2D Dirac Semimetal.}--We start with the effective tight-binding Hamiltonian $H_{\text{TB}}$ on a square lattice, whose realization shall be presented below:
\begin{eqnarray}
H_{\text{TB}} &=& \sum_{\bold k} \left( \begin{matrix}
c_{\bold k\uparrow}^\dagger, c_{\bold k\downarrow}^\dagger
\end{matrix}\right) \mathcal{H}_{\text{TB}}\left( \begin{matrix}
c_{\bold k\uparrow} \\
c_{\bold k\downarrow}
\end{matrix}\right); \nonumber \\
\mathcal{H}_{\text{TB}} &=& \left( m_z - 2t_x \cos k_x - 2t_z \cos k_z\right) \sigma_z \nonumber \\
&+&2t_{so} \sin k_z \sigma_y + t_{xI} \sin k_x \sigma_0 + m_x \sigma_x, \label{HTB}
\end{eqnarray}
where $c_{\bold k s}$ ($c^\dag_{\bold ks}$) is annihilation (creation) operator with spin $s=\uparrow,\downarrow$, $t_{x/z}$ is the hoping constant along $x/z$ direction, $t_{so}$ is the strength of spin-flip hopping, and $m_{z,x}$ denote the effective Zeeman couplings. As described in Fig.~\ref{Fig:Realization}(a), the above Hamiltonian describes a Dirac semimetal if $m_x=0$ and $|m_z|<2(t_x+t_z)$, with two Dirac points at $\mathbf{\Lambda_{\pm}} = \left(\pm \cos^{-1} [(m_z-2t_x)/2t_z],0\right)$. The term $t_{xI} \sin k_x \sigma_0$ breaks the inversion symmetry and leads to an energy difference between the two Dirac points. Finally, a nonzero $m_x$ opens a local gap at the two Dirac points. Without loss of generality, we take that $t_z = t_0, t_x = t_0 \cos \theta_0$ and $t_{xI} = 2 t_0 \sin \theta_0$ to facilitate the further discussion. As we show below, the above Hamiltonian can be readily realized in experiment.

The realization is sketched in Fig.~\ref{Fig:Realization}(b,c). The ingredients of realization include a blue-detuned spin-independent square lattice and a Raman lattice generated via only a single Raman transition (details can be found in the Supplementary Material~\cite{SI}). In particular, the two standing-wave lights (red and black lines in Fig.~\ref{Fig:Realization}(b)) of frequency $\omega_1$ form the electric field components $\bold E_{1x}=2E_1\hat e_z\cos k_0x$ and $\bold E_{1z}=2E_1(\alpha\hat e_x\cos k_0z+\beta\hat e_y\sin k_0z)$, and generate the square lattice potentials $V_{0x}\cos^2 k_0x$ and $V_{0z}\cos^2k_0z$, respectively, with $V_{0x}\propto|E_1|^2$ and $V_{0z}\propto(\alpha^2-\beta^2)|E_1|^2$. For our purpose we take that $\alpha$ is large compared with $\beta$, namely, the field $\bold E_{1z}$ is mainly polarized in the $\hat e_x$ direction. The $\pi/2$-phase difference between $\hat e_x$ and $\hat e_y$ polarized components is easily achieved by putting a $\lambda/4$-wave plate before mirror $M_1$, as shown in Fig.~\ref{Fig:Realization}(b). All the initial phases of the lights are irrelevant and have been neglected. To generate the Raman lattice another running light of frequency $\omega_2$ is applied along a tilted direction so that its wave vector $\bold k_2=k_0(\hat{x}\cos\theta+\hat{y}\sin\theta)$ (blue line in Fig.~\ref{Fig:Realization} (b)). $\bold E_{2}$ together with the light $\bold E_{1z}$ induces the Raman transition between spin-up $|\uparrow\rangle=|9/2,9/2\rangle$ and spin-down $|\downarrow\rangle=|9/2,7/2\rangle$, as illustrated in Fig.~\ref{Fig:Realization}(c). The generated Raman potential takes the form $M_{\rm eff}=2M_0(\alpha \cos k_0z + \beta \sin k_0z)e^{ik_1x}$$|$$\uparrow\rangle\langle\downarrow$$|$, with $k_1=k_0\cos\theta$ and $M_0\propto|E_1E_2|$. The former $\alpha$-term in $M_{\rm eff}$ leads to a spin-flip hopping along $z$ direction $t^{\rm so}_{\vec j,\vec j\pm1_z}=\pm(-1)^{j_z} e^{i\pi j_x\cos\theta}t_{\rm so}$, and the latter $\beta$-term gives an onsite Zeeman term $m_{\vec j}\sigma_x$, with $m_{\vec j}=(-1)^{j_z}e^{i\pi j_x\cos\theta}m_x$~\cite{SI}. Note that a small $\beta$-term can induce a relatively large $m_x$ compared with hopping couplings, since $m_x$ is induced by on-site coupling. Through a gauge transformation $c_{\vec j,\downarrow}\rightarrow c_{\vec j,\downarrow}(-1)^{j_z}e^{i\pi j_x\cos\theta}$~\cite{Liu2014}, we finally get the effective Hamiltonian~\eqref{HTB}, with the parameters being $t_x=t_0\cos[(\pi/2)\cos\theta]$ and $t_{xI}=2t_0\sin[(\pi/2)\cos\theta]$. It can be seen that the inversion symmetry is controlled by the tilt angle $\theta$, and the gap opening at Dirac points is controlled by the $\beta$, the $\hat e_y$-component of the $\bold E_{1z}$ field.

The above realization is clearly simpler than the recent experiments on 2D SO coupling, for it involves only one Raman transition. We note that a 2D Dirac semimetal driven by SO interaction, being different from graphene whose Dirac points are protected by symmetry only when there is no SO coupling, has not been discovered in solid state materials~\cite{Kane2015}. The high feasibility and controllability ensure that the 2D Dirac semimetal proposed here can be well observed based on the current experiment.

\begin{figure}[h]
\centerline{\includegraphics[width=1\columnwidth]{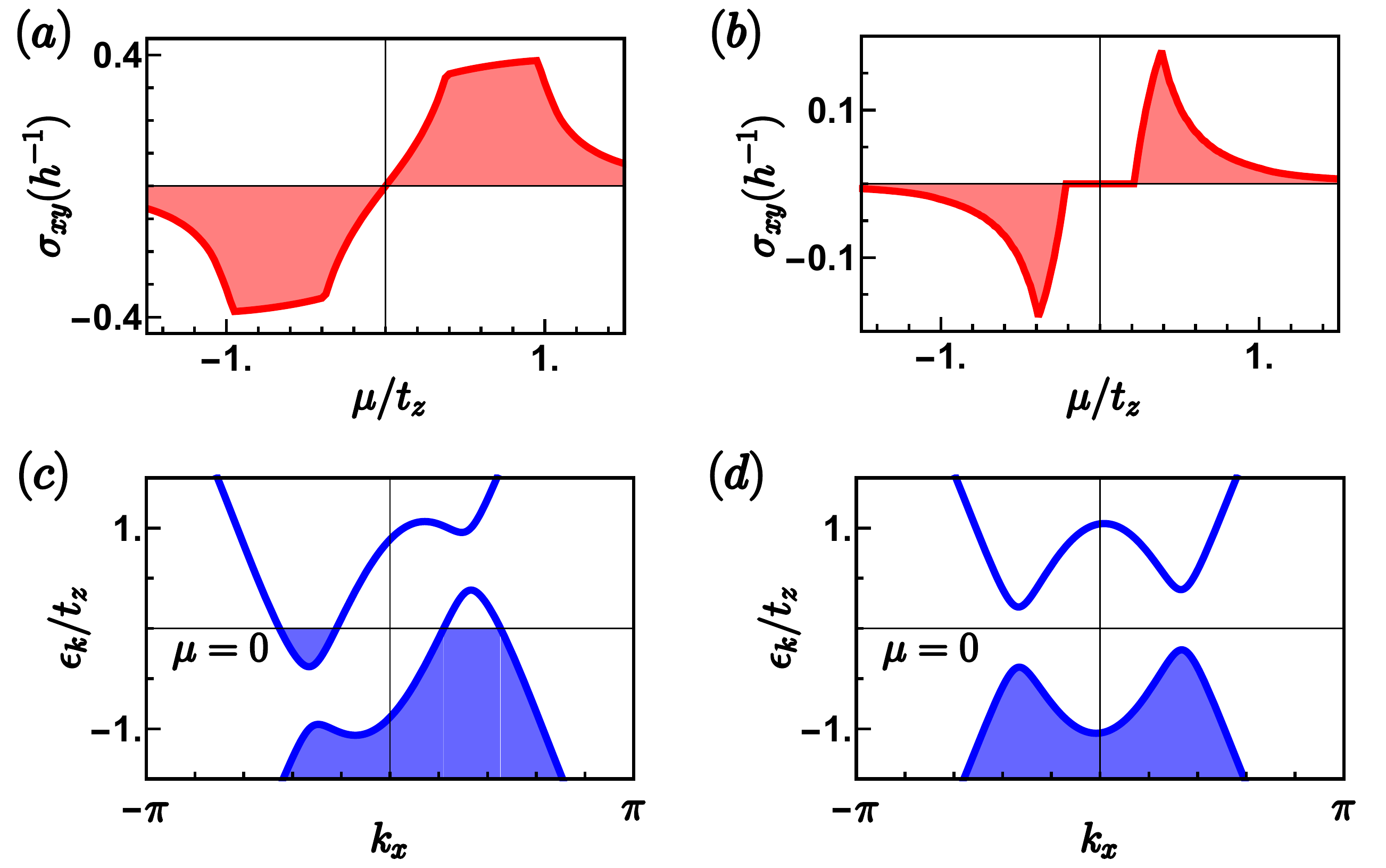}}
\caption{(a)-(b) The Hall conductaces $\sigma_{xy}$ versus chemical potential for the band structures (c) and (d) at $k_z=0$. In (b) an insulating regime can be reached. The parameters are rescaled by taking $t_z=1$, and we take $m_z=3, t_{so}=1, m_x = 0.3$ and $t_x =0.92$, where $t_{xI}=0.8t_x$ for (a) and (c), and $t_{xI}=0.1t_z$ for (b) and (d).}
\label{Fig:AHE}
\end{figure}
\textit{Anomalous Hall effect.}--When both the inversion and time-reversal symmetries are broken, the Hamiltonian~\eqref{HTB} leads to anomalous Hall effect, which reflects the nontriviality of 2D SO interaction~\cite{AHE}. Fig.~\ref{Fig:AHE} shows the hall conductance $\sigma_{xz}$ versus chemical potential $\mu$ and the spectra in different regimes. The hall conductance at zero temperature is calculated by ($\Theta$ is the step function)
\begin{eqnarray}
\sigma_{xz} &=& \frac{2}{h} \int dk_x dk_z \sum_{n,m} \Theta\left(\epsilon_m-\mu\right) \left(1-\Theta\left(\epsilon_n-\mu\right)\right) \nonumber \\
&&\times\text{Im}\left[\frac{\left(\left<u_n\left|\frac{\partial\mathcal{H}}{\partial k_z}\right|u_m\right>\left<u_m\left|\frac{\partial\mathcal{H}}{\partial k_x}\right|u_n\right>\right)}{\left(\epsilon_n-\epsilon_m\right)^2}\right]. \label{HallCon}
\end{eqnarray}
It can be seen that $\sigma_{xz}$ is always zero at $\mu=0$ or when $\mu$ is in the band gap [Fig.~\ref{Fig:AHE}(b,d)]. However, the nonzero $\sigma_{xz}$ for finite $\mu$ crossing bulk bands shows the nontrivial 2D SO effect, implying that the superfluid phase of nontrivial topology may be obtained.

\textit{States of superfluidity.}--
The superfluid phase can be induced by considering an attractive Hubbard interaction. The total Hamiltonian $H=H_{\rm TB}-U\sum_i n_{i\uparrow} n_{i \downarrow}$. Due to the existence of multiple FSs corresponding to different Dirac cones, in general we can have intra-cone pairing (FFLO) and inter-cone (BCS) pairing orders, defined by $\Delta_{2q} = \left(U/N\right) \sum_k \langle c^\dagger_{q+k\uparrow} c_{q-k\downarrow}\rangle$, with $q=\pm Q$ or $0$~\cite{Chan2016,Yao2015,Burkov2015,Wang2016}. On the other hand, since the inversion symmetry is broken, the BCS pairing would be typically suppressed. The topology of the superfluid phase can be characterized by the Chern number.

However, to compute the Chern number of the present system is highly nontrivial. The reason is because the FFLO order breaks translational symmetry can fold up the original Brillouin zone into many sub-Brillouin zones. In particular, if $Q=p\pi/q$, with $p$ and $q$ being mutual prime integers, the system includes $q$ sub-Brillouin zones. Moreover, if $Q$ is incommensurate with original lattice, the system folds up into infinite number of sub-Brillouiin zones. In such generic way, it is not realistic to compute numerically the Chern number of the system.

\begin{figure}[h]
\centerline{\includegraphics[width=\columnwidth]{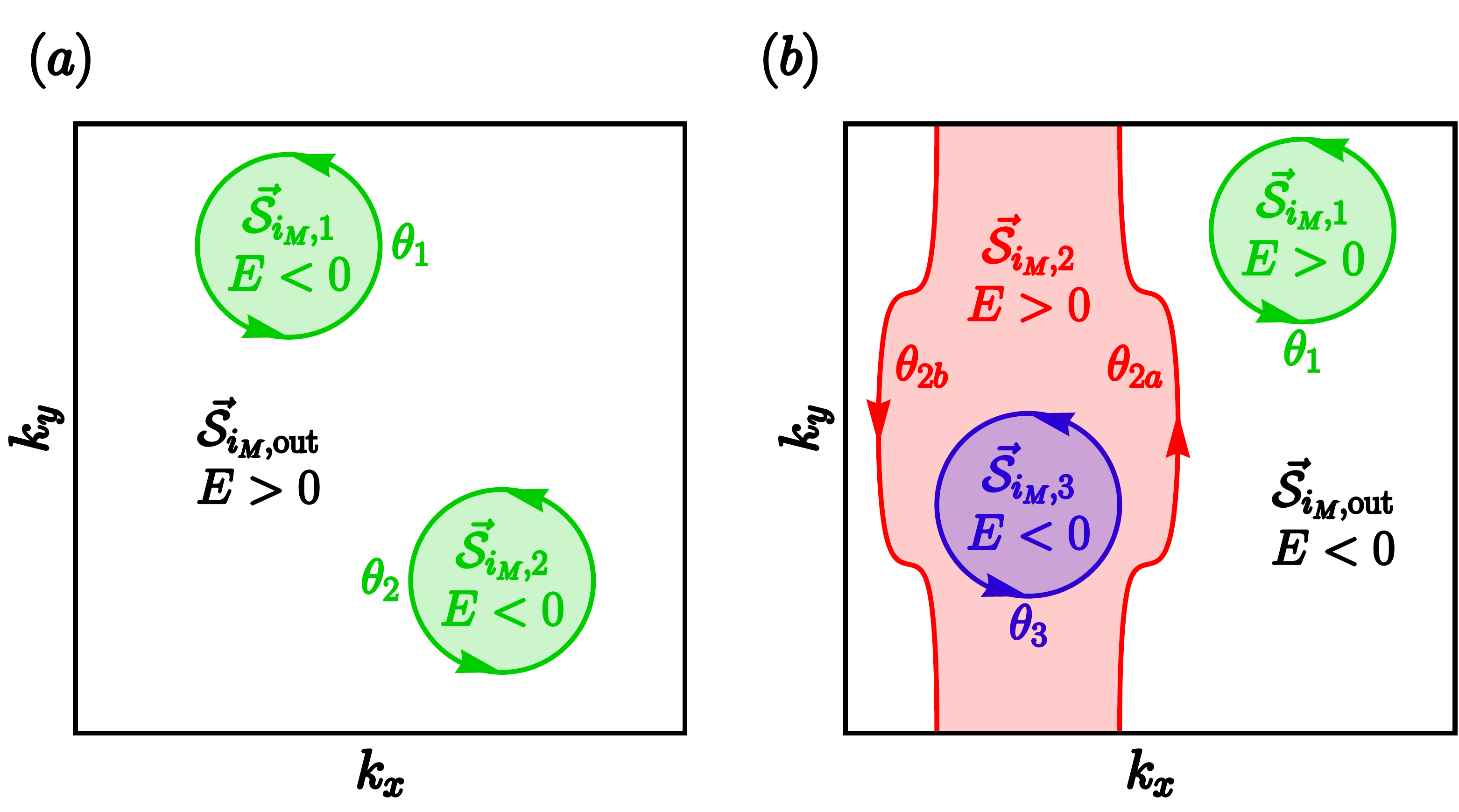}}
\caption{(a)-(b) Configurations of FSs of normal bands. The solid lines with arrows denote the FSs, which are boundaries of the vector areas $\vec S_{i_M,j}$ in the $\bold k$ space. The blue and green colors represent the closed FSs, and the red color represents open FSs.}
\label{Fig:GenT}
\end{figure}

\textit{Generic theory for Chern number.}--To resolve the difficulty, we show a generic theorem to determine the Chern number of a 2D system after opening a gap through having superfluid/superconductivity. For convenience we classify the normal bands of the system without pairing into three groups: upper (lower) bands which are above (below) the Fermi energy, and the middle bands which are crossed by Fermi energy. Each middle band may have multiple FSs (loops), and we denote by $(i_M,j)$ the $j$-th FS loop of the $i_M$-th middle band. Let the total Chern number of the upper (lower) bands be $n_U$ ($n_L$). We can show that the Chern number the superfluid pairing phase induced in the system is given by
\begin{eqnarray}\label{Chern1}
{\rm Ch}_1 &=& n_L - n_U + \sum_{i_M}\biggr[(-1)^{q_{i_M}}n_F^{(i_M)}- \nonumber \\
&&\sum_j (-1)^{q_{i_M,j}}\oint_{\partial \mathcal{\vec S}_{i_M,j}} \nabla_{\bold k} \theta_{\bold k}^{(i_M,j)} \cdot d{\bold k} \biggr]. \label{eqn:generalGenT}
\end{eqnarray}
Here $n_F^{(i_M)}$ is the Chern number of the $i_M$-th middle band and $\theta_{\bold k}^{(i_M,j)}={\rm arg}[\Delta^{(i_M,j)}_{\bold k}]$ is the phase of the pairing order projected onto the $(i_M,j)$-th FS loop~\cite{note1}. The integral direction is specified by arrows along FS lines in Fig.~\ref{Fig:GenT} (a,b), which defines the boundary $\partial \mathcal{\vec S}_{i_M,j}$ of the vector area $\vec S_{i_M,j}$ in $\bold k$ space. The quantities $\{q_{i_m,j},q_{i_M}\}=\{0,1\}$ are then determined by ``right-hand rule" specified below. The quantity $q_{i_M,j}=0$ (or $1$) if the energy of normal states within the area $\mathcal{\vec S}_{i_M,j}$ is positive (or negative), while $q_{i_M}=1$ (or $0$) if the regions ($\vec S_{i_M,\rm out}$) unenclosed by any FS loop have positive (or negative) energy (Fig.~\ref{Fig:GenT}).
With this theorem the Chern number of the superfluid phase can be simply determined once we know the properties of lower and upper bands, and the normal states at the Fermi energy, which govern the phase of $\Delta^{(i_M,j)}_{\bold k}$.

To facilitate the presentation, here we show the above theorem for a multiband system, but with only a single FS. The generic proof can be found in the Supplementary Material~\cite{SI}. We write down the BdG Hamiltonian by
\begin{eqnarray}
{\cal H}_{\text{\text{BdG}}}(\bold k)= \left(\begin{matrix}
{\cal H}(\bold k)-\mu & \Delta(\bold k) \\
\Delta^\dag(\bold k) & -{\cal H}^T(-\bold k)+\mu
\end{matrix}\right),
\end{eqnarray}
where ${\cal H}$ is the normal Hamiltonian, $\Delta(\bold k)$ is the pairing order matrix, and $\bold k$ is the local momentum measured from FS center. Note that if FS is not symmetric with respect to its center, one can continuously deform it to be symmetric without closing the gap. Finally we can always write down ${\cal H}_{\rm BdG}$ in the above form to study the topology. Denote by $u_{\bold k}^{(i_M)}$ the eigenvector of the normal band crossing Fermi energy. Focusing on the pairing on FS, the eiegenstates of ${\cal H}_{\rm BdG}$ has the form $[\alpha_k^{(i_M)}u_k^{(i_M)}, \beta_k^{(i_M)}u_{-k}^{(i_M)*}]^T$. The Chern number of the superfluid phase: ${\rm Ch}_1=n_L-n_U+\oint\nabla_{\bold k} \times \mathcal{A}_k^{(i_M)}d^2k$, with
\begin{eqnarray}\label{Berryconnection}
\mathcal{A}_{\bold k}^{(i_M)} &=& i [\alpha^{(i_M)}({\bold k}) u_{\bold k}^{(i_M)}]^\dag\nabla_{\bold k}\bigr[\alpha^{(i_M)}({\bold k}) u_{\bold k}^{(i_M)}\bigr]+\nonumber\\
&&i[\beta^{(i_M)}({\bold k}) u_{-{\bold k}}^{(i_M)*}]^\dag\nabla_{\bold k}\bigr[\beta^{(i_M)}({\bold k}) u_{-{\bold k}}^{(i_M)*}\bigr].
\end{eqnarray}
Let $\theta_k^{(i_M)}= {\rm arg}\bigr[\langle u_k^{(i_M)}|\Delta(k)| u_{-k}^{(i_M)*}\rangle\bigr]$ be the phase of order parameter on FS. Taking that $\Delta(k) \rightarrow \gamma \Delta(k)$ with $\gamma \rightarrow 0^+$ (without closing bulk gap), and with some algebra we can reach
\begin{eqnarray}
{\rm Ch}_1 &=&n_L-n_U +(-1)^{q_{i_M}} \biggr[n_F^{(i_M)}+2 \oint_{\partial{\vec S}_{i_M}} \widetilde{\mathcal{A}}_k^{(i_M)} \cdot dk \nonumber\\
& -&2\int_{\mathcal{\vec S}_{i_M}} \nabla_{\bold k}\times\widetilde{\mathcal{A}}_k^{(i_M)}d^2k - \oint_{\partial \mathcal{\vec S}_{i_M}} \nabla_k \theta_k^{(i_M)}\cdot dk \biggr] \label{eqn:gaugeGenT}
\end{eqnarray}
with $\widetilde{\mathcal{A}}^{(i_M)}_k=i [u_{\bold k}^{(i_M)}]^\dag\nabla_{\bold k}u_{\bold k}^{(i_M)}$ being the Berry connection for the normal band states. If we choose a gauge so that $\widetilde{\mathcal{A}}^{(i_M)}_k$ is smooth on $\mathcal{\vec S}_{i_M}$, the two terms regarding $\widetilde{\mathcal{A}}^{(i_M)}_k$ in the right hand side of Eq.~\eqref{eqn:gaugeGenT} cancels. We then reach the formula~\eqref{Chern1} for the case with a single FS.

The proof can be generalized to the case with generic multiple FSs which may be closed or open~[Fig.~\ref{Fig:GenT}(b)], with multiple bands crossed by Fermi energy, and with the pairing within each FS or between two different FSs, given that the pairing fully gaps out the bulk~\cite{SI}. Since the result is not restricted by pairing types, the generic theorem shown here is powerful to quantitatively determine the topology of the superfluid phases. On the other hand, it should be noted that this theorem is applied to judge the topology of the phase gapped out by small pairing orders. Starting from the phase governed by Eq.~\eqref{Chern1}, when the magnitude of superfluid order increases, the system may undergo further topological phase transition and enter a new phase with different topology. Monitoring such phase transitions can determine the topological region of the phase diagram.

\textit{Phase Diagram.}--
With the above theorem we can easily determine the topology of the superfluid phase. Note that for the present Dirac system, a BCS pairing cannot fully gap out the bulk but leads to nodal phases, while the FFLO or FF order can. When there is only one FS, the system with an FF order is topological since $n_L=n_U=n_F^{(i_M)}=0$ and $\int_{\partial\vec S_{(i_M)}} \nabla \theta^{(i_M)}_k\cdot dk=\pm1$, giving the Chern number ${\rm Ch}_1=\mp1$. In contrast, when there are two FSs, from the same or different bands, one can readily find that the contributions from both FSs cancels out and the Chern number is zero, rendering a trivial phase. This result implies that the system can be topological when it is in an FF phase.

\begin{figure}[h]
\centerline{\includegraphics[width=\columnwidth]{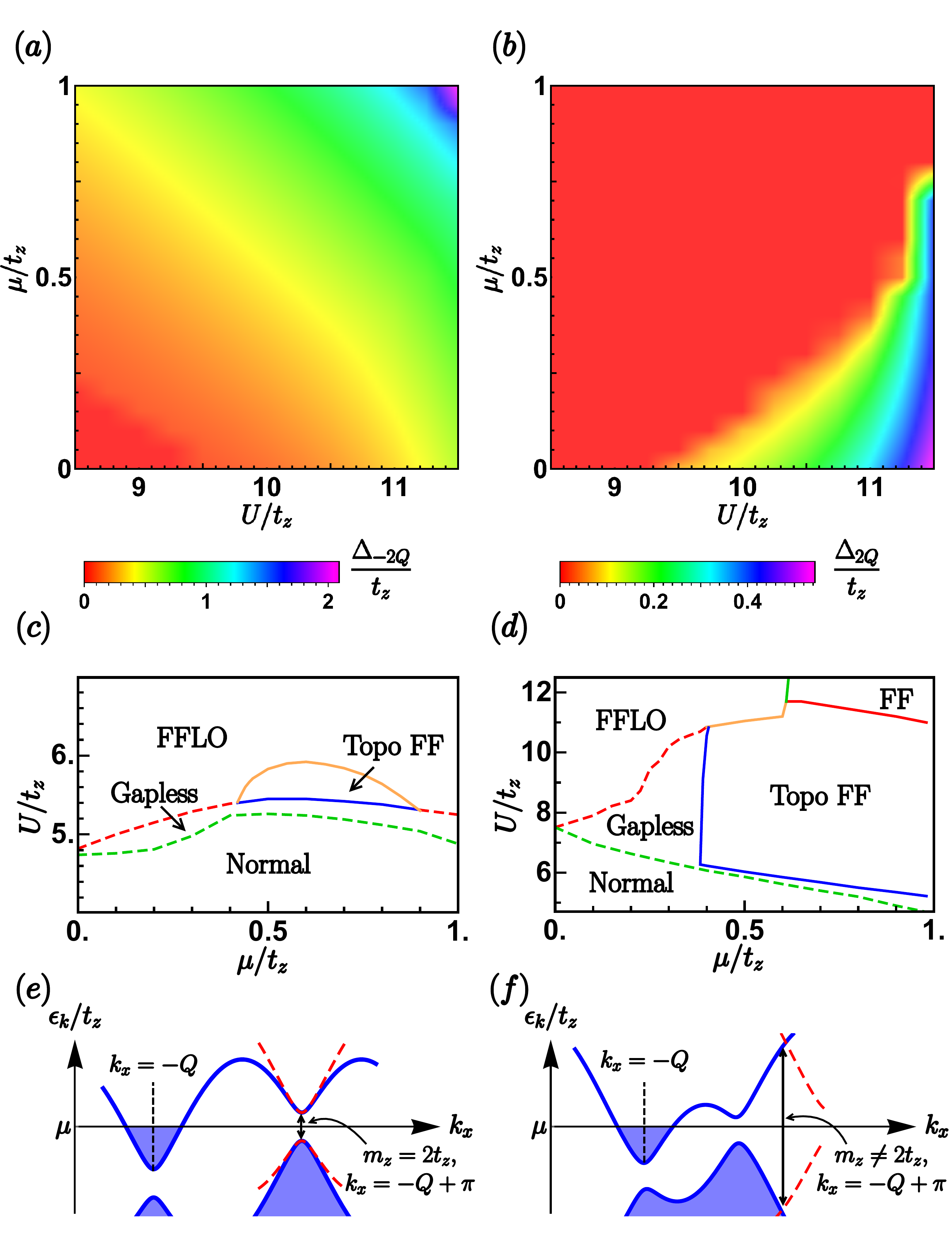}}
\caption{(a)-(b) The superfluid order $\Delta_{-2Q}$ dominates over others for $\mu>0$. (c) Phase diagram with normal-size topological region when $t_x=t_{so}=t_z, t_{xI}=0.7t_z, m_z=2t_z, m_x=0.3t_z$. (d) Phase diagram with broad topological region. The values of the parameters in diagrams (a,b,d) are $m_z=2.92t_z, t_x=0.92t_z, t_{so}=t_z, t_{xI}=0.8t_z, m_x=0.3tz$. Schematic diagrams showing the underlying mechanism for the normal-size (c) and broad (d) topological regions.}
\label{DeltaCal}
\end{figure}

From the mean field results shown in Fig.~\ref{DeltaCal} (a,b), we can see that the BCS pairing is greatly suppressed, and $\Delta_{-2Q}$ dominates over $\Delta_{2Q}$, with $\pm Q\approx\bold\Lambda_\pm$ for positive $\mu$. A rich phase diagram is given in Fig.~\ref{DeltaCal} (c,d), where the topological and trivial FF phases, and FFLO phase are obtained. It is particularly interesting that in Fig.~\ref{DeltaCal} (d) a broad topological region is predicted when $m_z$ is away from $m_z=2t_z$. The broad topological region implies that the upper critical value $\Delta_{2q}^{(c)}$, characterizing the transition from topological FF state to other phases, is largely enhanced compared with the case for $m_z=2t_z$. This is a novel effect explained below. Note that the superfluid order $\Delta_{-2Q}$ also couples the particle-hole states at $\tilde{\bold Q}=(\pi-Q,0)$. Increasing $\Delta_{-2Q}$ to $\Delta_{-2Q}^{(c)}$ closes the bulk gap at $\tilde{\bold Q}$ momentum, with the critical value being solved from BdG Hamiltonian as
\begin{equation}
\Delta_{-2Q}^{(c)}=\sqrt{m_x^2+m_p^2-\left(\frac{t_{xI}}{t_x}\sqrt{t_x^2-\frac{m_p^2}{16}}-\mu\right)^2},
\end{equation}
where $m_p = 2(m_z-2t_z)$. For $m_z=2t_z$, we have $Q=\pi/2$ and a small critical value $\Delta_{-2Q}^{(c)}\lesssim m_x$. This is because the gap closes at the right hand Dirac point $\tilde{\bold Q}=Q$, where the original bulk gap less than $2m_x$ before having superfluid pairing [Fig.~\ref{DeltaCal}(e)]. Importantly, for $m_z=3t_z$, we find that $\Delta_{-2Q}^{(c)} \sim 2t_z$, which is of the order of band width. In this regime $\tilde{\bold Q}$ is away from the right hand Dirac point, and corresponds to a relatively large bulk gap before adding $\Delta_{-2Q}$ [Fig.~\ref{DeltaCal}(f)]. As a result, a large $\Delta_{-2Q}$ is necessary to drive the phase transition, giving a broad topological region, as shown in Fig.~\ref{DeltaCal} (d).

\textit{Conclusion.}--In conclusion, we have proposed an experimental scheme to realize chiral Fulde-Ferrell (FF) superfluids, and showed a generic theorem to determine the topology of TSF phases through the normal states. We start with a tunable Dirac semimetal driven by 2D SO coupling, which has not been discovered in condensed matter materials but can be readily realized with the current ultracold atom experiments, and show how a superfluid phase with nonzero center of mass momentum is typically favored in the system.
Moreover, we have shown a generic formalism for the Chern number of a 2D superfluid induced in the normal states, with which the topological chiral FF superfluid with a broad topological region is predicted. Our findings are of significance for both ultracold atoms and condensed matter physics, and also can be useful for material science in search for new topological superconducting states.

This work is supported by MOST (Grant No. 2016YFA0301604), NSFC (No. 11574008), and Thousand-Young-Talent Program of China.

\onecolumngrid

\renewcommand{\thesection}{S-\arabic{section}}
\setcounter{section}{0}  
\renewcommand{\theequation}{S\arabic{equation}}
\setcounter{equation}{0}  
\renewcommand{\thefigure}{S\arabic{figure}}
\setcounter{figure}{0}  

\indent

\section*{\large Supplemental Material:\\
From semimetal to chiral Fulde-Ferrell superfluids}

\indent

In this Supplementary Material we provide the details of the model realization, and the proof of the generic theorem to determine the topology, i.e. the Chern number of a chiral superfluid/superconductor.

\section{Effective Hamiltonian}

\begin{figure}[h]
\centerline{\includegraphics[width=\columnwidth]{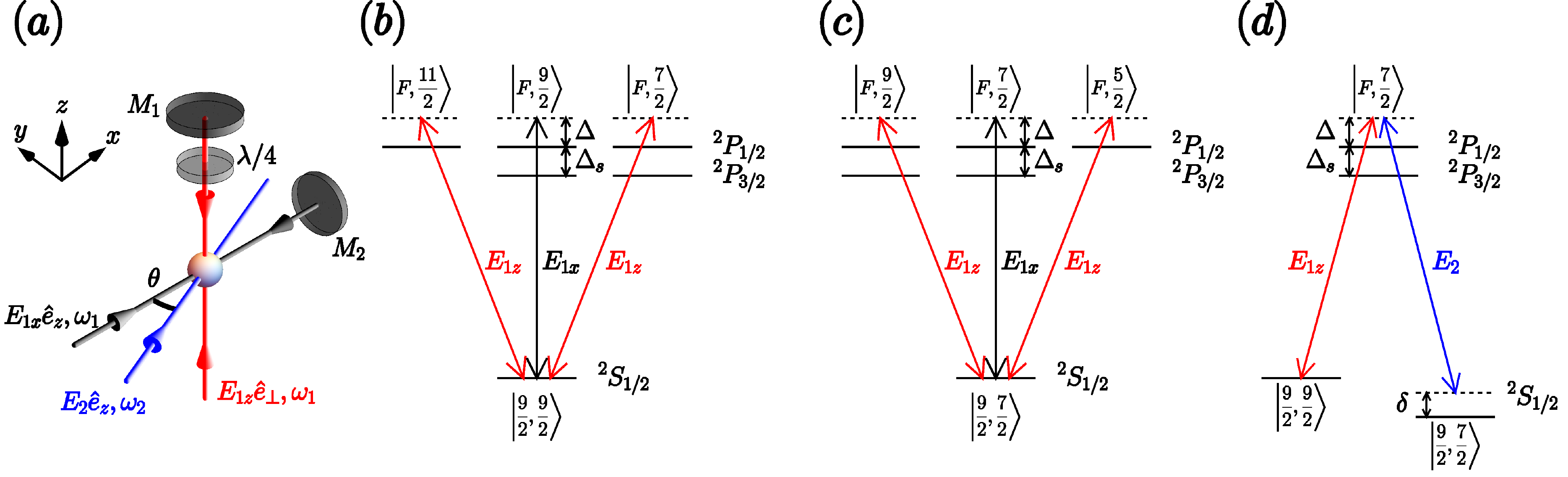}}
\caption{(a) Proposed experimental setting of realization. The standing wave lights formed by $\mathbf{E}_{1x,1z}$ generate a blue-detuned square lattice. The incident polarization of $\mathbf{E}_{1z}$ is $\hat{e}_\bot = \alpha \hat{e}_x + i \beta \hat{e}_y$, and the $\lambda/4$-wave plate changes the polarization of the reflected field to $\hat{e}'_\bot = \alpha \hat{e}_x - i \beta \hat{e}_y$. (b,c) As illustrated for $\hspace*{0pt}^{40}K$ fermions, the lattice potential for the state $\left| F=9/2, m_F=9/2  (7/2) \right>$ is generated by $\mathbf{E}_{1x,1z}$. (d) The Raman coupling is generated by $\mathbf{E}_{1z}$ and $\mathbf{E}_2$, which has a tilt angle $\theta$ with respect to $x$-axis in the $x$-$y$ plane.}
\label{Realization}
\end{figure}

\subsection{Atomic States}
The spin-1/2 pseudospin is defined by the two atomic states $\left| g_\uparrow \right> = \left| F=9/2 , m_F = 9/2 \right>$ and  $\left| g_\downarrow \right> = \left| F=9/2 , m_F = 7/2 \right>$ of the atom $\hspace*{0pt}^{40}K$. They are then coupled to the excited states in the manifold $\hspace*{0pt}^2P_{3/2}$ and $\hspace*{0pt}^2P_{1/2}$ through various two-photon processes depicted in Fig.~\ref{Realization}(b-d). \\

\subsection{Light fields}
The electric field of standing-wave lights for the present realization can be written as:
\begin{eqnarray}
\mathbf{E} &=& E_{1x} \left(e^{i\left(k_0 x + \phi_1\right)} + e^{i\left(-k_0 x + \phi_2\right)}\right)\hat{e}_z + E_{1z} \left(\alpha \left( e^{i\left(k_0 z + \phi_3\right)} + e^{i\left(-k_0 z + \phi_4\right)} \right)\hat{e}_x + \beta \left(e^{i\left(k_0 z + \phi_3\right)} + e^{i\left(-k_0 z + \phi_4 + \pi\right)}\right) \hat{e}_y \right) \nonumber \\
&&+ E_2 e^{i \left(k_1 y + \phi_5\right)} \hat{e}_x \nonumber \\
&=& 2 E_{1x}e^{i\left(\frac{\phi_1+\phi_2}{2}\right)}\cos \left(k_0 x + \frac{\phi_1-\phi_2}{2}\right)\hat{e}_z + 2E_{1z} \left(\alpha e^{i\left(\frac{\phi_3+\phi_4}{2}\right)}\cos \left(k_0 z + \frac{\phi_3-\phi_4}{2}\right)\hat{e}_x \right.\nonumber \\
&&+ \left. i\beta e^{i\left(\frac{\phi_3+\phi_4}{2}\right)}\sin \left(k_0 z + \frac{\phi_3-\phi_4}{2}\right)\hat{e}_y \right) + E_2 e^{i\left(k_1 y + \phi_5\right)} \hat{e}_x \nonumber \\
&=& 2e^{i\phi_B}E_{1x} \cos k_0 x \hat{e}_z + 2E_{1z} \left( \alpha \cos k_0 z \hat{e}_x + i\beta \sin k_0 z \hat{e}_y\right) + E_2 e^{i\left(k_1y + \phi_A\right)} \hat{e}_x,
\end{eqnarray}
where $\phi_A = -\frac{k_1}{k_0}\frac{\phi_1-\phi_2}{2}-\frac{\phi_3+\phi_4}{2}+\phi_5$, $\phi_B = \frac{\phi_1+\phi_2}{2}-\frac{\phi_3+\phi_4}{2}$. In the last line of the above equation we have made the change of parameters $x\rightarrow x-\left(\phi_1-\phi_2\right)/\left(2k_0\right)$ and $z\rightarrow z-\left(\phi_3-\phi_4\right)/\left(2k_0\right)$ and a multiplication of an overall phase factor $e^{-i\left(\phi_3+\phi_4\right)/2}$.

The Rabi-frequencies of the transitions described in Fig.~\ref{Realization}(b-d) can be derived. The light denoted by black line has the Rabi-frequency $\Omega_\pi = 2\Omega_\pi' E_{1x}\cos k_x$. Those lights depicted as red lines and connecting existing states with $\Delta F=\pm 1$ have the Rabi-frequency $\Omega_\pm = \frac{\Omega_\pm'}{\sqrt{2}} 2E_{1z}\left(\mp \alpha \cos k_0x + \beta \sin k_0z \right)$, and the one depicted as blue line is $\Omega_p = \Omega_p' E_2 e^{ik_0x\cos\theta} = \Omega_p' E_2 e^{ik_1x}$. The phase $\phi_{A,B}$ are irrelevant and so are omitted here. The quantities $\Omega_i'$ is the dipole matrix element $\left<g\left|\mathbf{r}_q\right|e\right>$, where $g$ and $e$ is the corresponding ground state and excited state, and $\mathbf{r}_q$ is the spherical component of vector $\mathbf{r}$.

\begin{table}
\begin{equation}
\begin{array}{|c|c|c|c|c|c|c|c|c|}
\cline{1-4}\cline{6-9}
\frac{1}{2} \rightarrow \frac{3}{2}; \frac{9}{2} & q=-1 & q=0 & q=1 & \hspace*{50pt}& \frac{1}{2} \rightarrow \frac{3}{2}; \frac{7}{2}
 & q=-1 & q=0 & q=1\\
\cline{1-4}\cline{6-9}
F'=\frac{11}{2} & \sqrt{\frac{1}{2}} & -\sqrt{\frac{1}{11}} & \sqrt{\frac{1}{110}} & &
F'=\frac{11}{2} & \sqrt{\frac{9}{22}} & -\sqrt{\frac{9}{55}} & \sqrt{\frac{3}{110}}\\
\cline{1-4}\cline{6-9}
F'=\frac{9}{2} &  & \sqrt{\frac{8}{33}} & -\sqrt{\frac{16}{297}} & & F'=\frac{9}{2} & \sqrt{\frac{16}{297}} & \sqrt{\frac{392}{2673}} & -\sqrt{\frac{256}{2673}}\\
\cline{1-4}\cline{6-9}
F'=\frac{7}{2} &  &  & -\sqrt{\frac{14}{135}} & &
F'=\frac{7}{2} &  & \sqrt{\frac{28}{1215}} & \sqrt{\frac{98}{1215}}\\
\cline{1-4}\cline{6-9}
\end{array} \nonumber
\end{equation}
\begin{equation}
\begin{array}{|c|c|c|c|c|c|c|c|c|}
\cline{1-4}\cline{6-9}
\frac{1}{2} \rightarrow \frac{1}{2}; \frac{9}{2} & q=-1 & q=0 & q=1 & \hspace*{50pt}& \frac{1}{2} \rightarrow \frac{1}{2}; \frac{7}{2}
 & q=-1 & q=0 & q=1\\
\cline{1-4}\cline{6-9}
F'=\frac{11}{2} &  & \sqrt{\frac{1}{3}} & -\sqrt{\frac{2}{27}} & &
F'=\frac{11}{2} & \sqrt{\frac{2}{27}} & \sqrt{\frac{49}{243}} & -\sqrt{\frac{32}{243}}\\
\cline{1-4}\cline{6-9}
F'=\frac{9}{2} &  &  & \sqrt{\frac{16}{27}} & & F'=\frac{9}{2} & & \sqrt{\frac{32}{243}} & \sqrt{\frac{112}{243}}\\
\cline{1-4}\cline{6-9}
\end{array} \nonumber
\end{equation}
\caption{The value of the ratio of $\left<g\left|\mathbf{r}_q\right|F',m_F'=m_F-q\right>$ and $\left|\left<J=1/2 \left|\left| e\mathbf{r} \right|\right| J' \right> \right| = \alpha_i $. The top-left corners of each table contains three numbers denoting $J$, $J'$ and $m_F$, respectively, where $J$ and $J'$ is the quantum number of $\mathbf{J}$ corresponding to the ground and excited states respectively and $\left|g\right> = \left|F=\frac{9}{2},m_F\right>$.} \label{DipoleT}
\end{table}

\subsection{Lattice and Raman potentials}

The lattice potential is generated by the two-photon processes which intermediate states are in the manifold $\hspace*{0pt}^2S_{1/2}$ (described in the diagram Fig.~\ref{Realization}(b,c) by the process generated by black and red lines). Each process gives a contribution $\sum_j \left|\Omega_i\right|^2/\Delta_j$ to the lattice potential, where $\Omega_i$ is the corresponding Rabi-frequency, $j$ runs through all possible atomic states and $\Delta_j = \Delta$ or $\Delta+\Delta_s$ if the intermediate state considered is in the manifold $\hspace*{0pt}^2P_{1/2}$ or $\hspace*{0pt}^2P_{3/2}$. According to experimental data, for reference, see ~\cite{TGK40}) all different $\Delta$ or $\Delta+\Delta_s$ corresponding to intermediate states of different $m_F$ are of the same order of magnitude (order of THz) and the differences among each group are negligible (order of 100MHz to 1GHz).

The lattice potential of both spin is
\begin{equation}
V\left(x,z\right) = \frac{4}{3}\left(\frac{2\alpha_2^2}{\Delta}+\frac{\alpha_1^2}{\Delta+\Delta_s}\right)\left(E_{1x}^2 \cos^2 k_0x+\left(\alpha^2-\beta^2\right) E_{1z}^2 \cos^2 k_0z\right),
\end{equation}
where we have omitted the constant term, and $\alpha_i = \left|\left< J=\frac{1}{2}\left|\left|e\mathbf{r}\right|\right|J=i-\frac{1}{2}\right>\right|$ ($i=1,2$) is the reduced matrix element between the total angular momentum $J=1/2$ and $J=3/2$. (The coefficients in $V_0$ can be calculated with the help of Table~\ref{DipoleT} in a straightforward way.) Notice that the coupling between spin-up $\left|9/2,9/2\right>$ and spin-down $\left|9/2,7/2\right>$ state through these processes are negligible since $E_{g_\uparrow} - E_{g_\downarrow} \gg \left| \Omega^2/\Delta \right|$, where $E_{g_s}$ is the energy of the spin-up or spin-down state.

The Raman lattice is generated via only one Raman process (described in the diagram Fig.~\ref{Realization}(d)). The Raman potential generated is given by the formula $\sum_{j}\Omega_-^* \Omega_p / \Delta_j$. In this case, the generated potential is
\begin{equation}
M_{\rm eff}=M_0\left(\alpha\cos k_0z+\beta \sin k_0z \right)e^{ik_1x}|\uparrow\rangle\langle\downarrow|, \label{RamanF}
\end{equation} where $M_0=\frac{1}{9}\left(\frac{\alpha_2^2}{\Delta}-\frac{\alpha_1^2}{\Delta+\Delta_s}\right)\sqrt{2}E_2E_{1z}$.
The Zeeman term $m_z \sigma_z$ is generated by the a small off-resonant in the Raman process, where $m_z = \tilde{\delta}/2 = \left(E_{g_\uparrow} -  E_{g_\downarrow} - \omega_1+\omega_2\right)/2$.

\subsection{Tight-binding Model}

We derive the tight-binding model by considering the hopping contributed from lattice and Raman potentials, respectively. The lattice potential contributes the spin-conserved hopping terms as
\begin{equation}
H_{\rm TB}^{(1)}=\left(-t_x' \sum_{j_x,j_z} C^\dagger_{j_x,j_z,\uparrow}C_{j_x+1,j_z,\uparrow}+C^\dagger_{j_x,j_z,\downarrow}C_{j_x+1,j_z,\downarrow}-t_z \sum_{j_x,j_z} C^\dagger_{j_x,j_z,\uparrow}C_{j_x,j_z+1,\uparrow}+C^\dagger_{j_x,j_z,\downarrow}C_{j_x,j_z+1,\downarrow}\right)+h.c.,
\end{equation}
where $t_x' = -V_0E_{1x}^2 \int dxdz \phi^*_{0,0}\left(x,z\right) \cos^2 k_0x \phi_{1,0}\left(x,z\right)$, $t_z = -V_0\left(\alpha^2-\beta^2\right)E_{1z}^2 \int dxdz \phi^*_{0,0}\left(x,z\right) \cos^2 k_0z \phi_{0,1}\left(x,z\right)$ and $k_0a=\pi$, and $\phi_{j_x,j_z}$ is the wavefunction at centered at $\left(x,z\right)=\left(j_x,j_z\right)a$.
On the other hand, for the spin-flip term induced by Raman coupling, we have that the first term of~\eqref{RamanF} provides spin-flip hopping along the $z$-direction of strength
\begin{eqnarray}
&&\alpha M_0\int dxdz \phi^*_{j_x,j_z}\left(x,z\right) \cos k_0z e^{ik_1x} \phi_{j_x,j_z\pm 1}\left(x,z\right) \nonumber \\
&=& \alpha M_0\int dxdz \phi^*_{0,0}\left(x,z\right) \cos k_0\left(z+j_za\right) e^{ik_1(x+j_xa)} \phi_{0,\pm 1}\left(x,z\right) \nonumber \\
&=& \left(-1\right)^{j_z}e^{i\frac{k_1}{k_0}\pi j_x} \alpha M_0\int dxdz \phi^*_{0,0}\left(x,z\right) \cos k_0z e^{ik_1x} \phi_{0,\pm 1}\left(x,z\right) \nonumber \\
&=& \mp \left(-1\right)^{j_z}e^{i\frac{k_1}{k_0}\pi j_x}  t_{so}, \nonumber
\end{eqnarray}
where $t_{so}=-\alpha M_0\int dxdz \phi^*_{0,0}\left(x,z\right) \cos k_0z e^{ik_1x} \phi_{0,1}\left(x,z\right) $. The same term gives no contribution to the hopping along $x$-direction since $\cos k_0z$ is antisymmetric in the $x$-direction at the local minimum of lattice potential.
Likewise, the second term of~(\ref{RamanF}) gives rise to an onsite Zeeman term
\begin{eqnarray}
&&\beta M_0\int dxdz \phi^*_{j_x,j_z}\left(x,z\right) \sin k_0z e^{ik_1x} \phi_{j_x,j_z}\left(x,z\right) \nonumber \\
&=&\beta M_0\int dxdz \phi^*_{0,0}\left(x,z\right) \sin k_0\left(z+j_za\right) e^{ik_1\left(x+j_xa\right)} \phi_{0,0}\left(x,z\right) \nonumber \\
&=&\left(-1\right)^{j_z}e^{i\frac{k_1}{k_0}\pi j_x}  m_x,\nonumber
\end{eqnarray}
where $m_x = \beta M_0 \int dxdz \phi^*_{0,0}\left(x,z\right) \sin k_0z e^{ik_1x} \phi_{0,0}\left(x,z\right)$. This term gives negligible contribution to the hopping term since $\beta \ll 1$.
Therefore, in the tight-binding model, the Raman potential contributes
\begin{eqnarray}
H_{\rm TB}^{(2)}=\sum_{j_x,j_z}\left(-1\right)^{j_z}e^{i\frac{k_1}{k_0}\pi j_x}\left[t_{so}\left(c^\dagger_{j_x,j_z,\uparrow}c_{j_x,j_z+1,\downarrow}-c^\dagger_{j_x,j_z,\uparrow}c_{j_x,j_z-1,\downarrow}\right) + m_x c^\dagger_{j_x,j_z,\uparrow}c_{j_x,j_z,\downarrow} \right]+h.c.
\end{eqnarray}

The total tight-binding Hamiltonian reads $H_{\rm TB}=H_{\rm TB}^{(1)}+H_{\rm TB}^{(2)}$, which can be simplified by applying the gauge transformation
\begin{equation}
c_{j_x,j_z,\uparrow/\downarrow} \rightarrow (-i)^{j_z+j_x} e^{\pm i\left(\frac{k_1\pi}{2k_0} j_x+\frac{\pi}{2} j_z\right)} c_{j_x,j_z,\uparrow/\downarrow}.
\end{equation}
With the Fourier transformation $c_{j_x,j_z\sigma} \rightarrow \frac{1}{\sqrt{N}} c_{k\sigma} e^{i k \cdot \left(j_x,j_z\right)a}$, we can finally get the Bloch Hamiltonian of the tight-binding model
\begin{eqnarray}
\mathcal{H}_{\text{TB}} &=& \left( m_z - 2t_z \cos k_z\right) \sigma_z +2t_{so} \sin k_z \sigma_y + m_x \sigma_x + 2t_x'\left(\begin{matrix}
\cos \left(k_x+\frac{\pi k_1}{2k_0}-\frac{\pi}{2}\right) & 0\\
0 & \cos \left(k_x-\frac{\pi k_1}{2k_0}-\frac{\pi}{2}\right)
\end{matrix}\right) \nonumber \\
&=& \left( m_z - 2t_x \cos k_x - 2t_z \cos k_z\right) \sigma_z +2t_{so} \sin k_z \sigma_y + t_{xI} \sin k_x \sigma_0 + m_x \sigma_x,
\end{eqnarray}
where $t_x = t_x' \sin \left( \pi \cos \theta/2 \right)$ and $t_{xI} = 2t_x' \cos \left( \pi \cos \theta/2 \right)$. All the parameters are independently tunable except that $t_x$ and $t_{xI}$ are related by $t_x^2 + t_{xI}^2/4=t_0^2$.

It can be seen that the inversion symmetry is controlled by the tilt angle $\theta$, and the gap opening at Dirac points is controlled by the $\beta$, the $\hat e_y$-component of the $\bold E_{1z}$ field. Note that $m_x$ is induced by onsite spin-flip transition. Thus a small $\beta$-term in Eq.~\eqref{RamanF} can generate a relatively large $m_x$. For our purpose, we shall consider a small $\beta$-term compared with $\alpha$-term. Thus the $\bold E_{1z}$ field is mainly polarized in the $\hat e_x$ direction.

\section{BdG Hamiltonian}

An attractive Hubbard interaction can be described effectively as $H_U=-U\sum_i n_{i\uparrow} n_{i\downarrow}$. We introduce three order parameters $\Delta_{\pm 2Q}$ and $\Delta_0$ when considering superconducting pairing, where $\Delta_{2q} = \left(-U/N\right) \sum_k  \left< c_{q+k}^\dagger c_{q-k}^\dagger \right>$ and $q=\pm Q$ or $0$. If only one of them is non-zero, the BdG Hamiltonian can be written as $H_{\text{BdG}} = \sum_k \Psi_k^\dagger \mathcal{H}_{\text{BdG}} \Psi_k /2$, where
\begin{equation}
\mathcal{H}_{\text{BdG}} = \left( \begin{matrix}
{H}_{\text{TB}}(k) & \Delta_{2q}\\
\Delta_{2q}^\dag & -{H}_{\text{TB}}^T(2q-k)
\end{matrix} \right)  \label{BdGNotFolded}
\end{equation}
and the basis of $\Psi_k$ is $\left(c_{k\uparrow}, c_{k\downarrow}, c_{2q-k\uparrow}^\dagger, c_{2q-k\downarrow}^\dagger\right)^T$. The pairing matrix $\Delta_{2q}$ is assumed to be real valued, which can be set via a change of overall phase factor of $c$. Here, $q$ is the center-of-mass momentum of the non-zero pairing.

If there are more than one of them is non-zero, we need to fold the Brillouin zone. Let $q=Q=m\pi/n$, where $m$ and $n$ are coprime integers. Then the density of the BdG Hamiltonian can be written as
\begin{equation}
\mathcal{H}_{\text{BdG}}=
\left(\begin{matrix}
H_0 & 0 & 0 & \cdots & 0 & \widetilde{\Delta}_0 & \widetilde{\Delta}_{-Q} & 0 & \cdots & \widetilde{\Delta}_{2Q} \\
0 & H_{2Q} & 0 & \cdots & 0 & \widetilde{\Delta}_Q & \widetilde{\Delta}_0 & \widetilde{\Delta}_{-2Q} & \cdots & 0 \\
0 & 0 & H_{4Q} & \cdots & 0 & 0 & \widetilde{\Delta}_Q & \widetilde{\Delta}_0 & \cdots & 0 \\
\vdots & \vdots & \vdots & \ddots & \vdots & \vdots & \vdots & \vdots & \ddots & \vdots\\
0 & 0 & 0 & \cdots  &  H_{-2Q} & \widetilde{\Delta}_{-2Q} & 0 & 0 & \cdots & \widetilde{\Delta}_0\\
\widetilde{\Delta}_0^\dagger & \widetilde{\Delta}_Q^\dagger & 0 & \cdots & \widetilde{\Delta}_{-2Q}^\dagger & H'_{0} & 0 & 0 & \cdots & 0\\
\widetilde{\Delta}_{-2Q}^\dagger & \widetilde{\Delta}_0^\dagger & \widetilde{\Delta}_Q^\dagger & \cdots & 0 & 0 &  H'_{-2Q} & 0 & \cdots & 0\\
0 &\widetilde{\Delta}_{-2Q}^\dagger & \widetilde{\Delta}_0^\dagger & \cdots & 0 & 0 & 0 & H'_{-4Q} & \cdots & 0\\
\vdots & \vdots & \vdots & \ddots & \vdots & \vdots & \vdots & \vdots & \ddots & \vdots\\
\widetilde{\Delta}_{2Q}^\dagger & 0 & 0 & \cdots  &  \widetilde{\Delta}_0^\dagger & 0 & 0 & 0 & \cdots & H'_{2Q}\\
\end{matrix}\right) \label{BdGFolded},
\end{equation}
where $H^{(')}_{rQ} = \pm H^{\left(T\right)}\left(\pm k_x+rQ,\pm k_z\right)$, $\widetilde{\Delta}_{2pQ} = \left(\begin{matrix}0 & \Delta_{2pQ} \\-\Delta_{2pQ} & 0\end{matrix}\right)$, $r = 0,1,2,\cdots,n-1$ and $p=0 \text{ or} \pm 1$.

The Nambu basis in the folded Brillouin zone, where $k_y \in \left[0,2\pi/n\right]$, is
\begin{equation}
\left(\begin{matrix}
c_{k_x,k_y} & c_{k_x+2Q,k_z} & \cdots & c_{k_x-2Q,k_z} &
c_{-k_x,-k_z}^\dagger & c_{-k_x-2Q,-k_z}^\dagger & \cdots & c_{-k_x+2Q,-k_z}^\dagger
\end{matrix}\right)^T \nonumber.
\end{equation}
In this basis, the gap equation can be derived through diagonalizing~(\ref{BdGNotFolded}) or~(\ref{BdGFolded}) and then calculate the expectation values of the order parameters. Therefore, from a particular value of $U$ the order parameters can be solved self-consistently.

\section{BKT Transition}
\begin{figure}[h]
\centerline{\includegraphics[width=.4\columnwidth]{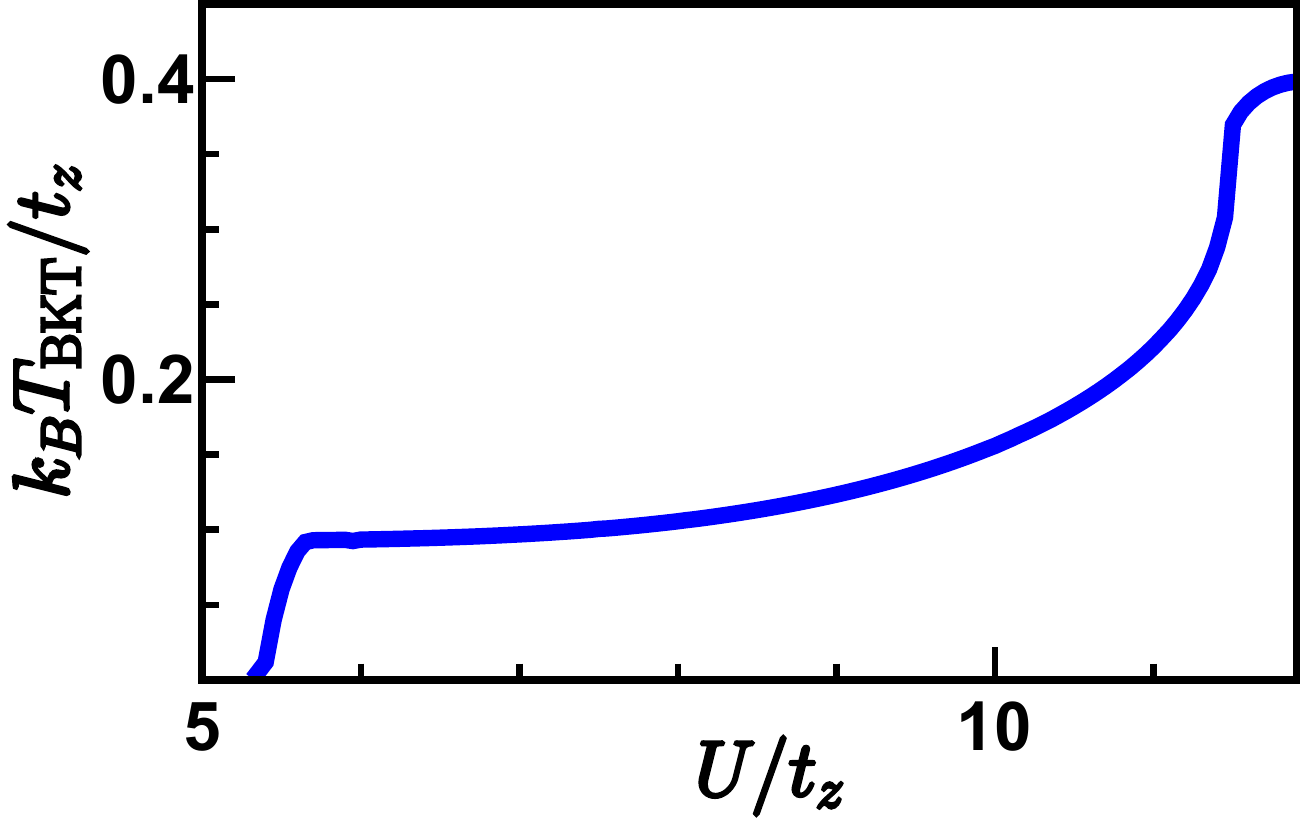}}
\caption{The BKT temperature when $m_z=2.92t_z, t_x=0.92t_z, t_{so}=t_z, t_{xI}=0.8t_z, m_x=0.3t_z$.}
\label{Fig:TBKT}
\end{figure}

The BKT Temperature can be calculated through the approximate BKT criterion
\begin{equation}
k_B T_{\text{BKT}} = \frac{\pi}{2} \rho_s \left(T_{\text{BKT}}\right) \approx \frac{\pi}{2} \rho_s \left(T=0\right),
\end{equation}
where $\rho_s$ is the superfluid density, which is can be calculated by $\rho_s=j_s/\delta q$. $j_s$ is the supercurrent density and $-Q$ is the center-of-mass momentum of two paired fermions when the system cut only one FS, which is of center-of-mass momentum $-Q$. $j_s$ is to be calculated by the variation of the zero-temperature energy, i.e. $j_s=\delta E_{\text{Total}}/\delta q$. The total energy $E_{\text{Total}}$ can be calculated through diagonalization of the total Hamiltonian in the Nambu basis, by assuming the center-of-mass momentum of the pairing to be $-Q+\delta q$. The numerical result is shown in Fig.~\ref{Fig:TBKT}

\section{Generic Theory for chiral topological superfluid}

After presenting the generic result, we shall show step by step the generic theory that the topology of 2D chiral superfluid/superconductor can be precisely determined by knowing only the properties of the Fermi surfaces and the topology of other normal bands away from the Fermi surfaces.

\subsection{Generic formalism}

For simplicity, we do not consider the case that the normal bands barely touch the Fermi energy. We classify the normal bands into three groups: upper (lower) bands, which are those with energy above (below) Fermi energy, and middle bands which are crossed by Fermi energy. Let $n_{L}$ ($n_U$) be the total Chern number of the upper (lower) band and $n_F^{(i_M)}$ be the Chern number of $i_M$-th middle band. Each middle band may form multiples Fermi surfaces. We denote the pairing of the $(i_M, j)$-th Fermi surface, projected on $i_M$-th band, to be $\Delta^{(i_M,j)}_k$ and its phase $\theta_k^{(i_M,j)} = \arg \Delta^{(i_M,j)}_k$.  We denote $\vec{\mathcal{S}}_{i_M,j}$ to be the area enclosed by the Fermi surface $\partial\vec{\mathcal{S}}_{i_M,j}$ and $\vec{\mathcal{S}}_{i_M,\text{out}}$ to be the unenclosed area. Notice that among each of the regions $\vec{\mathcal{S}}_{i_M,j}$ or $\vec{\mathcal{S}}_{i_M,\text{out}}$, the energy of the $i_M$-th band is of the same sign. We shall show that in general, the formula for the Chern number of the negative energy states, after considering the superconducting pairing, can be written as
\begin{equation}
Ch_1 = n_L - n_U +\sum_{i_M}  \left( (-1)^{q_{i_M}} n_F^{(i_M)} + \sum_j (-1)^{q'_{i_M,j}}\int_{\partial \vec{\mathcal{S}}_{i_M,j}} \nabla_k \theta_k^{(i_M,j)} \cdot dk \right),
\end{equation}
where the integral in the bracket is the winding number of $\theta_k^{(i_M,j)}$. The integral is evaluated in the direction satisfying right hand rule. The phase factor $(-1)^{q'_{i_M,j}}$ and $(-1)^{q_{i_M}}$ is $1$ if the energy of the $i_M$-th band is negative in the region $\vec{\mathcal{S}}_{i_M,j}$ and $\vec{\mathcal{S}}_{i_M,\text{out}}$  respectively, and they are $-1$ if the corresponding energy is positive. We have chosen a gauge here such that the phase of the Berry connection of the $i_M$-th band is continuous throughout the regions $\vec{\mathcal{S}}_{i_M,j}$. The case with an alternative gauge can be found below. This quantity is non-zero if the system is topological.

\subsection{One band, One FS}
First, we consider the case where there is only one band in the system and there is only one FS. So the Berry curvature, and thus the Chern number, of the normal states is zero since we can choose a gauge of real eigenvectors where the Berry connection is identically zero.\\

We can calculate the Berry connection, after considering the superconducting pairing, by direct calculation over each region $\vec{\mathcal{S}}$. The density of the BdG Hamiltonian and the Berry connection can be written as
\begin{eqnarray}
\mathcal{H}_{\text{BdG}}(k) &=& \left(
\begin{matrix}
\epsilon_k-\mu & \Delta_k \\
\Delta^*_k & -\epsilon_{-k}+\mu
\end{matrix}
\right); \nonumber\\
\mathcal{A}_{k\pm} &=& i \left(
\begin{matrix}
\alpha_\pm^*(k) & \beta_\pm^*(k) \\
\end{matrix}\right)  \nabla_k \left(
\begin{matrix}
\alpha_\pm(k) \\
\beta_\pm(k)
\end{matrix}\right) \nonumber,
\end{eqnarray}
where the $\pm$ sign denotes upper (lower) band.

$\alpha$ and $\beta$ can be found by diagonalizing the $\mathcal{H}_{\text{BdG}}$:
\begin{eqnarray}
\alpha_{\pm}(k)&=&\frac{\frac{\epsilon_k + \epsilon_{-k}}{2}-\mu \pm \sqrt{\left| \Delta_k\right|^2+\left(\frac{\epsilon_k + \epsilon_{-k}}{2}-\mu \right)^2}}{N_{\pm}(k)}; \nonumber\\
\beta_{\pm}(k)&=&\frac{\Delta_k^*}{N_{\pm}(k)}; \nonumber\\
N_{\pm}(k) &=& \sqrt{\left| N_{\pm}(k) \alpha_{\pm}(k) \right|^2 + \left| N_{\pm}(k) \beta_{\pm}(k) \right|^2}. \nonumber
\end{eqnarray}
So the Berry connection of the lower band is:
\begin{equation}
\mathcal{A}_{k}=i\frac{\Delta_k \nabla_k \Delta_k^* - \Delta_k^* \nabla_k \Delta_k}{4\sqrt{\left( \frac{\epsilon_k + \epsilon_{-k}}{2}-\mu \right)^2 + \left| \Delta_k\right|^2}\left(\sqrt{\left( \frac{\epsilon_k + \epsilon_{-k}}{2}-\mu \right)^2 + \left| \Delta_k\right|^2} - \left(\frac{\epsilon_k + \epsilon_{-k}}{2}-\mu\right)\right)}\nonumber.
\end{equation}
Since the Chern number remains unchanged under all continuous changes, we can assume that $\Delta = \gamma \Delta$ and let $\gamma \rightarrow 0$. We can also continuously deform the system so that the FS is symmetric. In this case, the Berry curvature is
\begin{equation}
\mathcal{B}_{k} = \nabla_k \times \mathcal{A}_{k} = \nabla_k \times \left(\frac{i\theta_{\pm\vec{\mathcal{S}}} }{2}\frac{\Delta_k \nabla_k \Delta_k^{*} - \Delta_k^{*} \nabla_k \Delta_k}{\left| \Delta_k \right|^2} \right) \nonumber,
\end{equation}
where $\theta_{\pm \mathcal{S}}$ denotes a function which valued to 1 inside (outside) the region $\vec{\mathcal{S}}$ and 0 outside (inside), and the upper (lower) sign means that the region $\vec{\mathcal{S}}$ is of positive (negative) energy. The Chern number can be calculated through an integral of $\mathcal{B}_{k}$
\begin{equation}
Ch_1 = \int \mathcal{B}_{k} d^2k = \mp \int_{\partial \vec{\mathcal{S}}} \nabla_k \theta_k \cdot dk, \label{SIChern1}
\end{equation}
where $\theta_k=\arg\Delta_k$ and that the Berry curvature localizes on the boundaries of $\vec{\mathcal{S}}$.

\subsection{One band, multiple FS} \label{Sec:OMFS}

To generalize the equation~(\ref{SIChern1}), we notice that when $\gamma \rightarrow 0$, the momentum modulation is unimportant. Thus the Chern number of the lower band depends only on the pairing within each FS. Therefore, the total Chern number is the summation of $\mp n^{(j)}$, where $n^{(j)}$ is the winding number of $\theta_k^{(j)} =\arg\Delta_k^{(j)}$, and $\Delta_k^{(j)}$ is the pairing order of the $j$-th FS; When the energy of the normal states is positive (negative) within the region $\vec{\mathcal{S}_j}$, we take the negative (positive) sign in $\mp n^{(j)}$.

\begin{figure}[h]
\centerline{\includegraphics[width=.9\columnwidth]{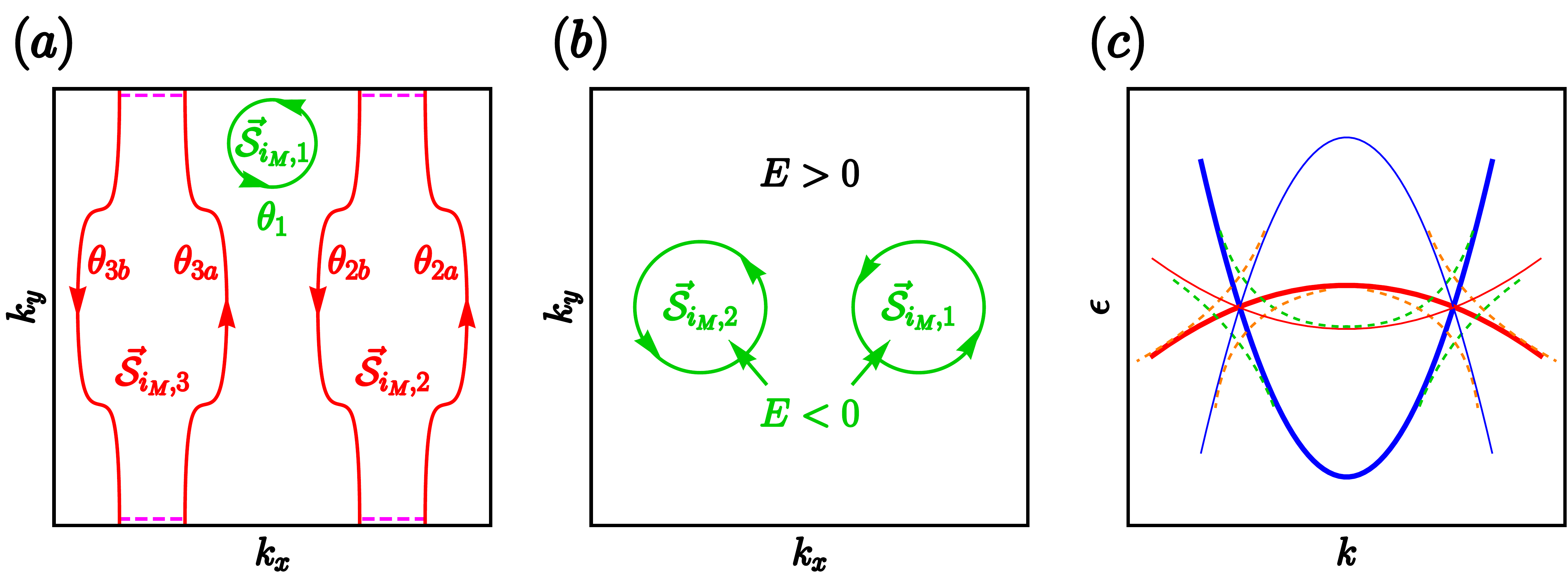}}
\caption{(a) Schematic diagram showing how the winding number should be counted. (b) Schematic diagram of the description in the proof concerning multiple FS in one band. (c) Schematic diagram of the case when two FS coincide with each another in folded Brillouin zone. Blue and red means the part of two bands which form the FS. Thick and thin means particle and its hole duplicates. The green and orange dashed line indicates how the band structure will change when the pairing between the bands denoted as blue and red lines are introduced.}
\label{GenTS}
\end{figure}

It is worthwhile to note that in general there maybe order parameters connecting two different FS (like the BCS order $\Delta_0$ in the main text). To simplify our discussion, we assume that the system contains minimal order parameters to be fully gapped. Without lost of generality, we can assume that the FS looks like Fig.~\ref{GenTS}(b). (If we have $E<0$ inside $\vec{\mathcal{S}}$, we can just flip the sign as indicated by~(\ref{SIChern1}).) Folding up the Brillouin zone, we will find that two FS coincides with each other and that
\begin{equation}
\left( \begin{matrix}
\epsilon_{k-Q} & 0 & \Delta_0\left(k-Q\right) & \Delta_{-Q}\left(k\right) \\
0 & \epsilon_{k+Q} & \Delta_Q\left(k\right) & \Delta_0\left(k-Q\right) \\
\Delta_0^*\left(k-Q\right)  & \Delta_{Q}^*\left(k\right) & -\epsilon_{-k+Q} & 0 \\
\Delta_{-Q}^*\left(k\right) & \Delta_0^*\left(k+Q\right)  & 0 & -\epsilon_{-k-Q} \\
\end{matrix} \right) \label{FBZChern}
\end{equation}
is the subspace of the BdG Hamiltonian in the folded Brillouin zone, where $\pm Q$ is the two center-of-mass momentum of the FS and $k$ is defined from $-Q/n$ to $Q/n$, where $n$ is an integer number. The minimal order parameters to fully gap the system can be $\Delta_{\pm Q}$, which we have discussed, or only $\Delta_0$ alone. When only $\Delta_0$ is non-zero, the winding numbers are calculated for $\Delta_0$ around $Q$ and $-Q$ (as indicated by the diagonal term in the $\Delta$ matrix).
\subsection{Multiple bands, One FS}
If there are many bands in the system, we call the Chern number of the upper (lower) band be $n_{L(U)}$. These bands contributes $n_L-n_U$ to the total Chern number for negative energy states in the duplicated space.

The arguments in the one band system need generalization since the Berry curvature of the normal bands are non-zero and an intuitive generalization of $\Delta_k^{(i_M)}$ to $\left<u_k^{(i_M)}\right| \Delta(k) \left|u_k^{(i_M)} \right>$, where $\left|u_k^{(i_M)} \right>$ is the eigenvector of the single particle Hamiltonian, is now gauge dependent.

The single particle Hamiltonian density $\mathcal{H}$ have eigenvectors $u_k^{(i_M)}$ so that $\mathcal{H}(k) u_k^{(i_M)} = \epsilon_k^{(i_M)} u_k^{(i_M)}$ and the BdG Hamiltonian density can be written as
\begin{equation}
\mathcal{H}_{\text{BdG}}(k) = \left(
\begin{matrix}
\mathcal{H}(k)-\mu & \Delta(k) \\
\Delta^\dag(k) & -\mathcal{H}^T(-k)+\mu
\end{matrix}
\right). \nonumber
\end{equation}
Write the BdG Hamiltonian density in the bases $
\left(\begin{matrix}
u_k^{(i)} & 0
\end{matrix}\right)^T$ and $
\left(\begin{matrix}
0 & u_{-k}^{(i)*}
\end{matrix}\right)^T$, $\mathcal{H}_\text{BdG}$ has diagonal entries with values $\epsilon_k^{(i_M)}-\mu$ and $-\epsilon_{-k}^{(i_M)}+\mu$ and the off-diagonal entries due to $\Delta(k)$, i.e. $\Delta_k^{(i,j)} = \left<u_k^{(i)}\right| \Delta(k) \left| u_{-k}^{(j)*}\right>$. Let $\Delta = \gamma \Delta$ and $\gamma \rightarrow 0$, the upper and lower bands are diagonalized automatically. The Hamiltonian in the subspace of middle bands can also be diagonalized as follows
\begin{eqnarray}
H_{\text{BdG}}(k)  \left(
\begin{matrix}
\alpha_\pm^{(i_M)}(k) u_k^{(i_M)} \\
 \beta_\pm^{(i_M)}(k) u_{-k}^{(i_M)*}
\end{matrix}\right) &=& \left(\frac{\epsilon_k^{(i_M)} - \epsilon_{-k}^{(i_M)}}{2} \pm \sqrt{\left| \Delta_k^{(i_M)}\right|^2+\left(\frac{\epsilon_k^{(i_M)} + \epsilon_{-k}^{(i_M)}}{2}-\mu \right)^2}\right) \left(
\begin{matrix}
\alpha_\pm^{(i_M)}(k) u_k^{(i_M)} \\
 \beta_\pm^{(i_M)}(k) u_{-k}^{(i_M)*}
\end{matrix}\right); \nonumber\\
\alpha_{\pm}^{(i_M)}(k)&=&\frac{\frac{\epsilon_k^{(i_M)} + \epsilon_{-k}^{(i_M)}}{2}-\mu \pm \sqrt{\left| \Delta_k^{(i_M)}\right|^2+\left(\frac{\epsilon_k^{(i_M)} + \epsilon_{-k}^{(i_M)}}{2}-\mu \right)^2}}{N_{\pm}^{(i_M)}(k)}; \nonumber\\
\beta_{\pm}^{(i_M)}(k)&=&\frac{\Delta_k^{(i_M)*}}{N_{\pm}^{(i_M)}(k)}; \nonumber\\
N_{\pm}^{(i_M)}(k) &=& \sqrt{\left| N_{\pm}^{(i_M)}(k) \alpha_{\pm}^{(i_M)}(k) \right|^2 + \left| N_{\pm}^{(i_M)}(k) \beta_{\pm}^{(i_M)}(k) \right|^2}, \nonumber
\end{eqnarray}
where $\Delta_k^{(i_M)}=\Delta_k^{(i_M,i_M)}$.\\

Therefore, when the $i_M$-th middle band is duplicated and gapped by $\Delta(k)$, the Berry connection of the upper and lower band are
\begin{eqnarray}
\mathcal{A}^{(i_M)}_{k\pm} &=& i \left(
\begin{matrix}
\alpha_\pm^{(i_M)*}(k) u_k^{(i_M)\dagger} & \beta_\pm^{(i_M)*}(k) u_{-k}^{(i_M)T} \\
\end{matrix}\right)  \nabla_k \left(
\begin{matrix}
\alpha_\pm^{(i_M)}(k) u_k^{(i_M)} \\
\beta_\pm^{(i_M)}(k) u_{-k}^{(i_M)*}
\end{matrix}\right). \nonumber \\
&=& i \alpha_{\pm}^{(i_M)*} \nabla_k \alpha_{\pm}^{(i_M)} + i \left| \alpha_{\pm}^{(i_M)} \right|^2 u_k^{(i_M)\dagger} \nabla_k u_k^{(i_M)} + i \beta_{\pm}^{(i_M)*} \nabla_k \beta_{\pm}^{(i_M)} + i \left| \beta_{\pm}^{(i_M)} \right|^2 u_{-k}^{(i_M)T} \nabla_k u_{-k}^{(i_M)*} \nonumber\\
&=& i \left(\left| \alpha_{\pm}^{(i_M)}\right|^2 - \left| \beta_{\pm}^{(i_M)} \right|^2 \right) u_k^{(i_M)\dagger} \nabla_k u_k^{(i_M)} \nonumber \\
&& + i\frac{\Delta_k^{(i_M)} \nabla_k \Delta_k^{(i_M)*} - \Delta_k^{(i_M)*} \nabla_k \Delta_k^{(i_M)}}{4\sqrt{\left( \frac{\epsilon_k^{(i_M)} + \epsilon_{-k}^{(i_M)}}{2}-\mu \right)^2 + \left| \Delta_k^{(i_M)}\right|^2}\left(\sqrt{\left( \frac{\epsilon_k^{(i_M)} + \epsilon_{-k}^{(i_M)}}{2}-\mu \right)^2 + \left| \Delta_k^{(i_M)}\right|^2} \pm \left(\frac{\epsilon_k^{(i_M)} + \epsilon_{-k}^{(i_M)}}{2}-\mu\right)\right)}\nonumber
\end{eqnarray}

There are three contributions when we calculate the Berry curvature, which is the curl of this quantity: (i) The curvature from the normal band;
(ii) The extra curvature due to the mixing of two bands (from the derivative of $|\alpha|^2-|\beta|^2$); (iii) The curvature due to superconducting pairing.

After continuously deforming the FS so that it is symmetric, we can write the Berry curvature of the lower band to be
\begin{eqnarray}
\mathcal{B}^{(i_M)}_{k\pm} &=& \pm\left(\left(\theta_{-\vec{\mathcal{S}}}-\theta_{\vec{\mathcal{S}}}\right) \widetilde{\mathcal{B}}_k^{(i_M)} + \left( \nabla_k \left(\theta_{-\vec{\mathcal{S}}}-\theta_{\vec{\mathcal{S}}}\right) \right) \times \widetilde{\mathcal{A}}_k^{(i_M)} \right)  \nonumber \\
&+& \nabla_k \times \left(\frac{i\theta_{\pm\mathcal{S}} }{2}\frac{\Delta_k^{(i_M)} \nabla_k \Delta_k^{(i_M)*} - \Delta_k^{(i_M)*} \nabla_k \Delta_k^{(i_M)}}{\left| \Delta_k^{(i_M)} \right|^2} \right) \nonumber,
\end{eqnarray}
where the upper (lower) sign is taken when the $i_M$-th band has positive (negative) energy in the region $\vec{\mathcal{S}}$, $\theta_{\pm\vec{\mathcal{S}}}$ denotes a function which valued to 1 inside (outside) $\vec{\mathcal{S}}$ and 0 outside (inside) $\mathcal{S}$ and $\widetilde{\mathcal{A}}_k^{(i_M)}$ ($\widetilde{\mathcal{B}}_k^{(i_M)}$) denotes the Berry connection (curvature) of the corresponding normal band. \\

The integral of Berry curvature can be simplified to (writing $\theta_k^{(i_M)} = \arg \Delta_k^{(i_M)}$):
\begin{eqnarray}
\int \mathcal{B}^{(i_M)}_{k\pm} dk &=& \mp\left(2\int_{\vec{\mathcal{S}}} \widetilde{\mathcal{B}}_k^{(i_M)}d^2k-n_F^{(i_M)}-2 \int_{\partial\vec{\mathcal{S}}} \widetilde{\mathcal{A}}_k^{(i_M)} \cdot dk + \int_{\partial \vec{\mathcal{S}}} \frac{i}{2}\frac{\Delta_k^{(i_M)} \nabla_k \Delta_k^{(i_M)*} - \Delta_k^{(i_M)*} \nabla_k \Delta_k^{(i_M)}}{\left| \Delta_k^{(i_M)} \right|^2} \cdot dk \right)\nonumber \\
&=& \mp\left(2\int_{\vec{\mathcal{S}}} \widetilde{\mathcal{B}}_k^{(i_M)}d^2k-n_F^{(i_M)}-2 \int_{\partial\vec{\mathcal{S}}} \widetilde{\mathcal{A}}_k^{(i_M)} \cdot dk + \int_{\partial \vec{\mathcal{S}}} \nabla_k \theta_k^{(i_M)}\cdot dk \right) \nonumber,
\end{eqnarray}

Notice that the third and fourth terms are gauge dependent, and they all add up to a gauge independent value. We now specify the gauge to finish the calculation. We require that:
\begin{equation}
\int_{\partial\vec{\mathcal{S}}} \widetilde{\mathcal{A}}_k^{(i_M)} \cdot dk = \int_{\vec{\mathcal{S}}} \widetilde{\mathcal{B}}_k^{(i_M)}d^2k \nonumber.
\end{equation}
We call this gauge $\vec{\mathcal{S}}$ and for the corresponding eigenfunctions we call them $u_{k[\vec{\mathcal{S}}]}^{(i_M)}$. Then the phase of $\Delta_k^{(i_M)}$ can be defined unambiguously. (i.e. we define the matrix element in $\vec{\mathcal{S}}$ gauge: $\Delta_{k[\vec{\mathcal{S}}]}^{(i,j)} = \left<u_{k[\vec{\mathcal{S}}]}^{(i)}\right| \Delta(k) \left| u_{-k[\vec{\mathcal{S}}]}^{(j)*}\right>$). The gauge merely means that the eigenvectors $u_k^{(i_m)}$ are continuous within the region $\vec{\mathcal{S}}$.

If we have other choice of gauge, we count the number of singularity (with the sign determined by right hand rule) within $\vec{\mathcal{S}}$. Each singularity contributes to $-2$ extra winding of $\theta_k^{(i_M)}$ and is canceled by the subtraction of integral of Berry connection from $2\int_{\vec{\mathcal{S}}} \widetilde{\mathcal{B}}_k^{(i_M)}d^2k$. Therefore, we know that the winding in different gauge can be related as
\begin{equation}
\int_{\partial \vec{\mathcal{S}}} \nabla_k \theta_{k[\vec{\mathcal{S}}]}^{(i_M)} \cdot dk= \int_{\partial \vec{\mathcal{S}}} \nabla_k \theta_{k[
\mathcal{G}]}^{(i_M)} \cdot dk + 2n_{s,\vec{\mathcal{S}}[\mathcal{G}]},
\end{equation}
where $n_{s,\vec{\mathcal{S}}[\mathcal{G}]}$ is the number of singularity inside $\vec{\mathcal{S}}$ in gauge $\mathcal{G}$.

Therefore total Chern number of the negative energy states is
\begin{equation}
Ch_1=n_L - n_U \pm \left( n_F^{i_M} - \int_{\partial\vec{\mathcal{S}}} \nabla_k \theta_{k[\vec{\mathcal{S}}]}^{(i_M)} \cdot dk \right) = n_L - n_U \pm \left( n_F^{(i_M)} - \int_{\partial \vec{\mathcal{S}}} \nabla_k \theta_{k[\mathcal{G}]}^{(i_M)} \cdot dk - 2n_{s,\vec{\mathcal{S}}[\mathcal{G}]} \right) \label{MOChern},
\end{equation}
for a general gauge $\mathcal{G}$.

\subsection{Multiple bands, Multiple FS}

Note that each particular middle band $i_M$ can contribute to the Chern number by
\begin{equation}
n_M^{(i_M)} = (-1)^{q_{i_M}} n_F^{(i_M)} + (-1)^{q'_{i_M,j}} \left( \int_{\partial \vec{\mathcal{S}}_{i_M,j}} \nabla_k \theta_{k[\mathcal{G}]}^{(i_M,j)} + 2n_{s,\vec{\mathcal{S}}_{i_M,j}[\mathcal{G}]} \right),
\end{equation}
where the phase factor $(-1)^{q'_{i_M,j}}$ and $(-1)^{q_{i_M}}$ is $1$ if the energy of the band is negative in the region $\vec{\mathcal{S}}_{i_M,j}$ and $\vec{\mathcal{S}}_{i_M,\text{out}}$  respectively, and they are $-1$ if the corresponding energy is positive.

Therefore, the contribution from all bands can be written as
\begin{equation}
Ch_1=n_L - n_U +\sum_{i_M} n_M^{(i_M)}.
\end{equation}

It is worth noting that as in this case, one more complexity arises. As in~(\ref{FBZChern}), we consider order parameter that connect two FS:
\begin{equation}
\left( \begin{matrix}
\epsilon_{k-Q}^{(i_M)} & 0 & \Delta_0^{(i_M,j_M)}\left(k-Q\right) & \Delta_{-Q}^{(i_M,i_M)}\left(k\right) \\
0 & \epsilon_{k+Q}^{(j_M)} & \Delta_Q^{(j_M,j_M)}\left(k\right) & \Delta_0^{(j_M,i_M)}\left(k-Q\right) \\
\Delta_0^{(i_M,j_M)*}\left(k-Q\right)  & \Delta_{Q}^{(j_M,j_M)*}\left(k\right) & -\epsilon_{-k+Q}^{(j_M)} & 0 \\
\Delta_{-Q}^{(i_M,i_M)*}\left(k\right) & \Delta_0^{(j_M,i_M)*}\left(k+Q\right)  & 0 & -\epsilon_{-k-Q}^{(i_M)} \\
\end{matrix} \right).
\end{equation}
If $\epsilon_{k-Q}^{(i_M)}$ and $\epsilon_{k+Q}^{(j_M)}$ are of the same sign in the coinciding region in the folded Brillouin zone, the argument in \ref{Sec:OMFS} have no difficulties except the similar generalization to the solution of the problem of allotment of Chern number and the problem of gauge. If $\epsilon_{k-Q}^{(i_M)}$ and $\epsilon_{k+Q}^{(j_M)}$ are of opposite sign, the minimal order parameter set cannot be $\Delta_0$ alone (see Fig.~\ref{GenTS}(c) for illustration). This is just the case for the model discussed in the main text, where $\Delta_0$ alone cannot gap the system.

We state without proof here that, if only $\Delta_0$ and one of the $\Delta_{\pm Q}$ is non-zero, the system can still be fully gap. Without lost of generality, suppose $\Delta_{Q}$ is non-zero, the $j_M^{\text{th}}$ band connect with its hole partner directly and $i_M^{\text{th}}$ band connect through two intermediate steps. The Chern number of this kind of interaction can also be calculated. The direct pairing is straightforward, while for the indirect pairing (the one containing two intermediate steps), we can count the winding number of the effective $\Delta$. In this case, the effective pairing between $-\epsilon_{-k-Q}^{(i_M)}$ and $\epsilon_{k-Q}^{(i_M)}$ are proportional to the multiplication of $\Delta_{0}^{/(j_M,i_M)}$, $\Delta_{-Q}^{(j_M,j_M)*}$ and $\Delta_{0}^{(i_M,j_M)}$, therefore the winding of the effective $\Delta$ is the sum of the winding of these three quantities. The problem concerning gauge and the allotment of the original Chern number can also be similarly investigated.


\begin{thebibliography}{2}

\bibitem{Zhan2016} Z. Wu, L. Zhang, W. Sun, X.-T. Xu, B.-Z. Wang, S.-C. Ji, Y. Deng, S. Chen, X.-J. Liu, and J.-W. Pan, Science {\bf 354}, 83
(2016).

\bibitem{Zhang2016a} L. Huang, Z. Meng, P.Wang, P. Peng, S.-L. Zhang, L. Chen,
D. Li, Q. Zhou, and J. Zhang, Nat. Phys. {\bf 12}, 540 (2016).

\bibitem{Zhang2016b} Z. Meng, L. Huang, P. Peng, D. Li, L. Chen, Y. Xu,
C. Zhang, P. Wang, and J. Zhang, Phys. Rev. Lett. {\bf 117}, 235304 (2016).

\bibitem{Ruseckas2005} J. Ruseckas, G. Juzeli\"{u}nas, P. \"{O}hberg, M. Fleischhauer, Phys.
Rev. Lett. {\bf 95}, 010404 (2005).
\bibitem{Zoller2005} K. Osterloh, M. Baig, L. Santos, P. Zoller, M. Lewenstein, Phys.
Rev. Lett. {\bf 95}, 010403 (2005).
\bibitem{Liu2009} X.-J. Liu, M. F. Borunda, X. Liu, J. Sinova, Phys. Rev. Lett. {\bf 102},
046402 (2009).

\bibitem{Read2000} N. Read and D. Green, Phys. Rev. B 61, 10267 (2000).
\bibitem{Kitaev2001} A. Y. Kitaev, Physics-Uspekhi 44, 131 (2001).
\bibitem{Kitaev2003} A. Y. Kitaev, Ann. Phys. (Amsterdam) 303, 2 (2003).
\bibitem{Fu2008} L. Fu and C. L. Kane, Phys. Rev. Lett. 100, 096407 (2008).
\bibitem{Sau2010} J. D. Sau, R. M. Lutchyn, S. Tewari, and S. Das Sarma, Phys. Rev. Lett. 104, 040502 (2010).
\bibitem{Lutchyn2010} R. M. Lutchyn, J. D. Sau, and S. Das Sarma, Phys. Rev. Lett. 105, 077001 (2010).
\bibitem{Oreg2010} Y. Oreg, G. Refael, and F. von Oppen, Phys. Rev. Lett. 105, 177002 (2010).

\bibitem{Zhang2008} C. Zhang, S. Tewari, R. M. Lutchyn, and S. Das Sarma,
Phys. Rev. Lett. 101, 160401 (2008).
\bibitem{Sato2009} M. Sato, Y. Takahashi, and S. Fujimoto, Phys. Rev. Lett.
103, 020401 (2009).

\bibitem{Zhu2011} S.-L. Zhu, L. B. Shao, Z. D.Wang, and
L.-M. Duan, Phys. Rev. Lett. 106, 100404 (2011).
\bibitem{Chunlei2013} C. Qu, Z. Zheng, M. Gong, Y. Xu, L. Mao, X. Zou, G. Guo, and C. Zhang, Nature Comm. {\bf 4}, 2710 (2013).
\bibitem{Zhang2013} W. Zhang and W. Yi, Nature Comm. {\bf 4}, 2711 (2013).
\bibitem{Hu2014} Y. Cao, S. -H. Zou, X.-J. Liu, S. Yi, G.-L. Long, and H. Hu, Phys. Rev. Lett. 113, 115302 (2014)

\bibitem{Liu2014} X. -J. Liu, K. T. Law, and T. K. Ng, Phys. Rev. Lett. \textbf{112}, 086401 (2014); {\it ibid} {\bf 113}, 059901 (2014).
\bibitem{Chan2016} C. Chan and X. -J. Liu, arXiv:1611.08516.

\bibitem{Wilczek2009} F. Wilczek, Nat. Phys. 5, 614 (2009).
\bibitem{Alicea2012} J. Alicea, Reports on Progress in Physics 75, 076501 (2012).
\bibitem{Franz2013} M. Franz, Nat. Nano. 8, 149 (2013).
\bibitem{Elliott2015} S. R. Elliott and M. Franz, Rev. Mod. Phys. 87, 137 (2015).

\bibitem{FF1964} P. Fulde and R. A. Ferrell,
Phys. Rev. \textbf{135}, A550 (1964).
\bibitem{LO1964} A. I. Larkin and Y. N. Ovchinnikov,
Zh. Eksp. Teor. Fiz. \textbf{47}, 1136 (1964).

\bibitem{SI} See the Supplementary Material for details of the realizatoin, and the proof of the generic theorem for computing the Chern number. The reference~\cite{TGK40} has been included.

\bibitem{Kane2015} S. M. Young and C. L. Kane
Phys. Rev. Lett. {\bf 115}, 126803 (2015).

\bibitem{TGK40} T. G. Tiecke,
Feshbach resonances in ultracold mixtures of the fermionic quantum gases $\hspace*{0pt}^{\text{6}}$Li and $\hspace*{0pt}^{\text{40}}$K. PhD thesis, University of Amsterdam, 2009.

\bibitem{AHE} N. Nagaosa, J. Sinova, S. Onoda, A. H. MacDonald, and P. Ong,
Rev. Mod. Phys. {\bf 82}, 1539 (2010).

\bibitem{Yao2015} S.-K. Jian, Y.-F. Jiang, and H. Yao, Phys. Rev. Lett. 114, 237001 (2015).
\bibitem{Burkov2015} G. Bednik, A. A. Zyuzin, and A. A. Burkov, Phys. Rev. B 92, 035153 (2015).
\bibitem{Wang2016} T. Zhou, Y. Gao, and Z. D. Wang, Phys. Rev. B 93, 094517 (2016).

\bibitem{note1} The pairing can occur between two different Fermi surfaces, say between $(i_M,j)$ and $(i_M',j')$. In this case the Eq.~\eqref{Chern1} is still valid, but the integral will be performed on both Fermi surfaces at the same time, see~\cite{SI}.
\end{thebibliography}
\end{document}